\documentclass[a4paper,fleqn,usenatbib]{mnras}

\usepackage[T1]{fontenc}
\usepackage{ae,aecompl}

\usepackage{graphicx}	
\usepackage{amsmath}	
\usepackage{amssymb}	

\newcommand{\beq}{\begin{equation}}
\newcommand{\eeq}{\end{equation}}
\newcommand{\bea}{\begin{eqnarray}}
\newcommand{\eea}{\end{eqnarray}}
\newcommand{\mb}{\mathbf}

\newcommand{\lx}{\langle}
\newcommand{\rx}{\rangle}

\title[Rayleigh scattering in dense fluid helium]{Rayleigh scattering in dense fluid helium}

\author[Ren\'e D. Rohrmann]{Ren\'e D. Rohrmann
\\Instituto de Ciencias Astron\'omicas, de la Tierra y del Espacio
   (CONICET), Av. Espa\~na 1512 (sur),5400 San Juan, Argentina\\}

\date{Accepted XXX. Received YYY; in original form ZZZ}

\pubyear{2017}

\begin{document}
\label{firstpage}
\pagerange{\pageref{firstpage}--\pageref{lastpage}}
\maketitle

\begin{abstract}
Iglesias et al. (2002) showed that the Rayleigh scattering from helium atoms decreases by collective effects in the atmospheres of cool white dwarf stars. Their study is here extended to consider an accurate evaluation of the atomic polarizability and the density effects involved in the Rayleigh cross section over a wide density-temperature region.
The dynamic dipole polarizability of helium atoms in the ground state is determinated with the oscillator-strength distribution approach. 
The spectral density of oscillator strength considered includes most significant single and doubly excited transitions to discrete and continuum energies. Static and dynamic polarizability results are confronted with experiments and other theoretical evaluations shown a very good agreement. In addition, the refractive index of helium is evaluated with the Lorentz-Lorenz equation and shows a satisfactory agreement with the most recent experiments. The effect of spatial correlation of atoms on the Rayleigh scattering is calculated with Monte Carlo simulations and effective energy potentials that represent the particle interactions, covering fluid densities between 0.005 and a few g/cm$^3$ and temperatures between $1000$~K and $15000$~K. We provide analytical fits from which the Rayleigh cross section of fluid helium  can be easily calculated at wavelength $\lambda>505.35$~\AA. Collision-induced light scattering was estimated to be the dominant scattering process at densities greater than 1--2~g/cm$^3$ depending on the temperature.
\end{abstract}




\section{Introduction}
\label{intro}  

Rayleigh scattering is an important source of opacity and emission in the study of phenomena of radiative transfer and interpretations of astronomical observations. Its cross section is required in atmosphere models to solve the radiative transfer equation and reproduce the spectrum of stars and sub-stellar objects.
The influence of the Rayleigh scattering on the electromagnetic radiation can also be used for the interpretation of the geometrical configuration of eclipsing binary stars \citep{Is89}, determinations of temperature and the scaleheight of atmospheres in transiting extrasolar planets \citep{Le08,Be12,Ho12,Be16}, abundance measurements in the atmosphere of giant planets \citep{Ca74,Pa06,Be15}, 
as a powerful consistency test on recombination physics \citep{Yu01,Le13}, and to establish a pressure standard and a more accurate value of the Boltzmann constant \citep{Sc07,Pu16}. 

Helium is the second most abundant element in the Universe and the knowledge of its scattering cross section is the 
interest for spectrum analysis and structure modeling of 
compact objects, such as  white dwarfs, brown dwarfs and giant planets. Specifically, helium opacity plays an important role in the time cooling of hydrogen-deficient white dwarfs at advanced stages of the evolution \citep{Ca17}. In fact, the cooling of such objects is very sensitive to the atmosphere opacity, which can affect the calibration of white dwarfs as cosmological chronometers \citep{Ga16}. 

Fiften years ago, \citet{Ig02} showed that collective effects in the gas of dense helium atmospheres of cool white dwarfs may yield a drastic reduction of gas opacity. The reduction of the scattering cross section by density effects is present in a variety of condensed systems \citep{Ma07,Pl15}, and it is related to the spatial arrangement of atoms and the consequent interference pattern of the scattered radiation. This interference pattern is accounted by the dynamical structure function, which contains information on the fluid structure and atomic dynamics \citep{Ha06}. Collective effects as specified by the dynamical structure function has long been treated in the scattering of neutrons, high energy electrons, and X-rays by atomic systems \citep{vH54,St93,Sc05,Sch07}. The basis for the calculation of scattering cross sections is the time-dependent quantum perturbation theory. When applied to dense systems, this approach shows that the Rayleigh scattering of radiation is a second-order, two-photon process which depend on both the dipole polarizability of atoms and the structure factor of the fluid \citep{ck07}.

Because Rayleigh scattering is an important source of opacity in rich-helium atmospheres of white dwarfs, it results important take into account many-body effects on its cross section, in order to evaluate accurate spectrum and photometry of these fosil stars and provide boundary conditions for evolutionary models \citep{Ca17}. Unfortunately, the study of \citet{Ig02} did not cover a broad range of temperatures and densities as required in such stellar atmosphere calculations. 
On the other hand, because that investigation was mainly aimed to show the main effects of high gas density, Iglesias et al. adopted an simplified description of the atomic polarizability where the contribution from highly excited and continuum states was neglected. 

The purpose of the current paper is to present a comprehensive analysis of the Rayleigh scattering in dense fluid helium, and to provide results for 
application in radiative tranfer calculations over a wide range of densities and temperatures. For this aim, we calculate the dynamic dipole polarizability throught the spectral density of oscillator strengths, 
and develop Monte Carlo simulations to evaluate the density effects on the scattering cross section. Simulations are performed taken into account effective potentials to describe the atomic interactions. 
We report the use of two pair potentials \citep{RY86,Az95} which incorporate averaged many-body interactions. One of them \citep{Az95} is finally choosen in view of the good agreement with results of \citet{Mi09} based on path integral Monte Carlo and density functional molecular dynamics. In addition, the current dipole polarizability was employed for the calculation of the refractive index of the fluid helium as a function of density and wavelength.
Simulation data were also used to study the relative importance of collision-induced light scattering over the range of density considered.

Dipole polarizability of helium has been studied extensively through a variety of theoretic \citep{Da62a,Mi10} and experimental methods \citep{Pe96,Sc07}. We follow here a semi empirical approach based on the spectral oscillator-strength distribution, combined with experimental data of photoionization cross-sections and theoretical results for discrete transitions. This method had early applications for helium \citep{Wh33,Vi33,Da57}, but the use of latest available data had remained unexplored. As will be shown in the next sections, the good agreement obtained with a variety of experimental and theoretical results demostrates that this is a very reliable method.

The paper is organised as follows. In Section \ref{tools} we present the basic ingredients to evaluated the Rayleigh-scattering cross section in dense fluids.
Section \ref{s.pol}  describes the atomic polarizability calculation based on the spectral density of oscillator strength, and the evaluation of the refractive index. Section \ref{s.s} is devoted to the numerical calculations of the atom-atom correlation functions and their effects on the Rayleigh cross section. Finally, the conclusions are given in Section \ref{conc}.

\section{Framework}
\label{tools}

\begin{figure}  
        \includegraphics[width=\columnwidth]{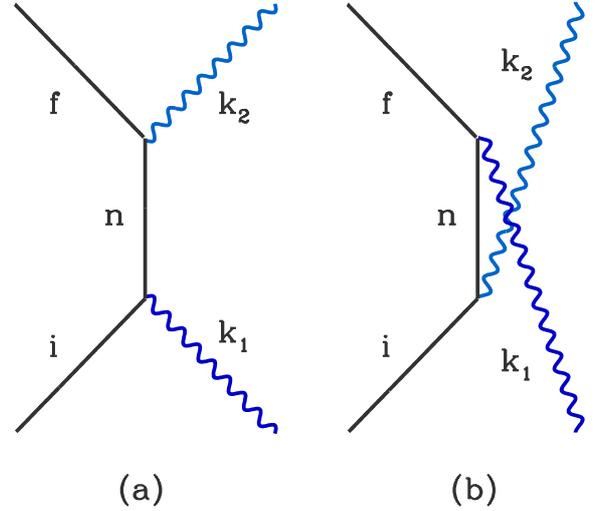}  
\caption{Diagrams for the scattering of light by an atom at second-order in the perturbative approach. Initial, intermediate, and final atomic states are denoted by $i$, $n$, and $f$, respectively; $k_1$ and $k_2$ specify the wavenumbers of the incoming and outgoing photons.}
\label{f.fdia}
\end{figure}

Light scattering consists on the absorption of a photon with wavenumber $\mb{k}_1$ (frequency $\omega_1$) and the emission of a photon with wavenumber $\mb{k}_2$ (frequency $\omega_2$), while the atom, initially in a state $|i\rx$, ends up in a state $|f\rx$. In the electric-dipole approximation and second order in the quantum pertubative approach \citep{Sa67, Lo00}, such two-photon process
demands the agency of intermediate atomic states $|n\rx$ into two pathways (virtual transitions) as shown in Fig. \ref{f.fdia}, where the photon $\mb{k}_1$ is annihilated first  $(a)$, or the photon $\mb{k}_2$ is created first $(b)$.
Rayleigh scattering corresponds to the case where the atom returns to the initial state ($|f\rx=|i\rx$), so it is nominaly an elastic or coherent process ($\omega_2=\omega_1$). However, in dense fluids, atoms are perturbated by neighbor ones and Rayleigh scattering may become slightly inelastic, while the associated cross section can be stronghly affected by density fluctuations in the fluid.

Collective effects on the scattering cross section $\sigma$ are accounted for the dynamical structure factor $S(\mb{k},\omega)$ \citep{vH54,Ha06}, where the wavenumber $\mb{k}$ and the frequency $\omega$ quantify the momentum ($\hbar \mb{k}$, with $\hbar=h/2\pi$, $h$ the Planck constant) and energy ($\hbar \omega$) tranferred to the atom in the scattering process, respectively,
\beq
\mb{k}=\mb{k}_2-\mb{k}_1,
\eeq
\beq
\omega=\omega_2-\omega_1.
\eeq
The function $S(\mb{k},\omega)$ represents the space and time Fourier transform of the density-density correlation function, which is the probability density of finding an atom at position $\mb{r}$ at time $t$ when there is one at $\mb{r}'$ at time $t'$.
As was shown by \citet{vH54}, the rate at which radiation is scattered by atoms into a solid angle $d\Omega$ and with outgoing frequency in the range $d\omega_2$, becomes proportional to $S(\mb{k},\omega)$. For radiation with wavelength bigger than atom size, and atomic state $|i\rx$ with spherical symmetry (such is the case of the helium ground state), the double differential cross-section of Rayleigh scattering is given by \citep{ck07}
\beq \label{sigma1}
\frac{d^2\sigma}{d\Omega d\omega_2}=  
\left( \frac{\omega_1}{c} \right)^4 \left|\mb{e}_1.\mb{e}^*_2\right|^2 \,
\alpha(\omega_1)^2 \,S(\mb{k},\omega),
\eeq
where $c$ is the light speed, $\mathbf{e}_1$ and $\mathbf{e}^*_2$ are the electric polarization of incoming and outgoing radiation, and $\alpha(\omega_1)$ is the dynamic dipole polarizability of atoms, which can be expressed as
\beq \label{alpha}
\alpha(\omega_1) =  \frac23 \sum_n  
\frac{(E_n-E_i)\left|\lx n|\mb{d}|i\rx \right|^2}{(E_n-E_i)^2-(\hbar\omega_1)^2},
\eeq
with $E_i$ and $E_n$ the eigenenergies associated to initial and intermediate atomic states, respectively, and $\mb{d}$ the atomic dipole moment. The summation in Eq. (\ref{alpha}) runs over all excited states which have electric dipole transitions with the ground state and, therefore, it means a summation over discrete states and an integration over the continuum spectrum.

Integration of (\ref{sigma1}) over frequency $\omega_2$ (for fixed incident frequency) yields the differential cross section
\beq \label{sigma2}
\frac{d\sigma}{d\Omega}= 
\left( \frac{\omega_1}{c} \right)^4 \left|\mathbf{e}_1.
\mathbf{e}^*_2\right|^2 \, \alpha(\omega_1)^2 \,S(\mb{k}),
\eeq
where 
\beq
S(\mb{k}) =\int S(\mb{k},\omega)   d\omega,
\eeq
is the static structure factor of the fluid.

\section{Atomic polarizability}  
\label{s.pol} 

The dynamic (static) polarizability expresses a quantitative measure of the distortion of a charge distribution (bound electrons at atoms in the present case) under the influence of an external monochromatic (steady) electric field. The dynamic dipole polarizability in atomic units 
[bohr$^3$] can be expressed in a Sellmeier form \citep{Se71,Da57,Bo97}
\beq \label{sellmeier}
\alpha(E)=4\left[ \sum_n \frac{f_n}{(E_n^2-E^2)}
+\int \frac{df/dE'}{(E'^2-E^2)} dE' \right],
\eeq
where all energies are measured in Rydbergs and atomic ones ($E_n$ and $E'$) are given respect that of the ground state ($1s^2\,^1S$ for helium). In Eq. (\ref{sellmeier}), $E$ is the incident radiation energy, $df/dE$ the differential oscillator strength for transitions from the ground state to the continuum energy, and $f_n$ the dipole oscillator strength for transitions to an excited state 
\beq
f_{n} =  \frac{2mE_n}{3e^2\hbar^2} \left|\lx n|\mb{d}|i\rx\right|^2,
\eeq
with $m$ and $e$ the electron mass and charge, respectively. The differential oscillator strength is related to the photoionization cross section $\sigma_\text{phot}(E)$ as follows \citep{Fc68}
\begin{equation} \label{dfdE}
\frac{df}{dE}=\frac{mc}{\pi e^2 h} \sigma_\text{phot}(E)
 =9.1107\times 10^{15}\frac{1}{\text{cm}^2\text{eV}}\sigma_\text{phot}(E).
\end{equation}
Because helium is a two-electron system, double as well as single electron jumps of both excitation and ionization transitions must be taking into account in evaluations of (\ref{sellmeier}).

\subsection{Energies and oscillator strenghts}  
\label{s.fosc}

Table \ref{t.fosc} contains wavelengths $\lambda_n$, energies $E_n$ and oscillator strengths $f_n$ of transitions in the $1^1S\rightarrow n^1P$ (He I) series taken from \citet{Th87} for $n\le21$, and \citet{Kh88} for $n=30$, $40$ and $50$. Energies for $n\ge 22$ have been calculated from
\begin{equation} \label{En}
  E_n = I_1-1.002866 \frac{I_\text{H}}{n^2},
\end{equation}
where $I_1=24.5876$~eV and $I_\text{H}=13.6057$~eV are the ionization energies of He and H atoms, respectively. Eq. (\ref{En}) is a fit to results compiled by \citet{Ma73} and yields single-excited state energies with an estimated error lower than $10^{-3}$~eV.
The $f$-values for highly excited states merge into the differential oscillator strength $df/dE$ in the continuous spectrum, according to \citep{Ha28,Ha29}
\begin{equation} \label{fnmerge}
\frac{n_*^3 f_n}{2R}=\left.\frac{df}{dE}\right|_{E=I_1},
\end{equation}
where $n_*$ is an effective main quantum number and $R$ is the Rydberg constant. In present evaluations, we assumed valid Eq. (\ref{fnmerge}) for transitions $1s^2~^1S\rightarrow 1sn^1P$  with $n_*=n\ge 22$, and used Eq. (\ref{dfdE}) with the photoionization head $\sigma(I_1)=7.40$~Mbarn measured by \citet{Sa94}.

%
\begin{table}
        \centering
        \caption{Oscillator strengths $f_n$ for the (He) $1^1S\rightarrow n^1P$ series taken from \citet{Th87} for $n\le21$ and \citet{Kh88} for $n=30$, 40 and 50, with $n$ the main quantum number, $E_n$ the transition energy and $\lambda_n$ the line wavelength. Numbers in brackets indicate power of ten.}
       \label{t.fosc}
       \begin{tabular}{rccc} 
       \hline
 $n$ & $\lambda_n$[\AA] & $f_n$ & $E_n$[eV] \\
       \hline
 2 & $584.3342$ & $2.7643(-1)$ & $21.2180$ \\
 3 & $537.0297$ & $7.3336(-2)$ & $23.0870$ \\
 4 & $522.2130$ & $2.9804(-2)$ & $23.7421$ \\
 5 & $515.6166$ & $1.4995(-2)$ & $24.0458$ \\
 6 & $512.0983$ & $8.6030(-3)$ & $24.2110$ \\
 7 & $509.9980$ & $5.3897(-3)$ & $24.3107$ \\
 8 & $508.6431$ & $3.5999(-3)$ & $24.3755$ \\
 9 & $507.7179$ & $2.5233(-3)$ & $24.4199$ \\
10 & $507.0578$ & $1.8356(-3)$ & $24.4517$ \\
11 & $506.5704$ & $1.3779(-3)$ & $24.4752$ \\
12 & $506.2002$ & $1.0607(-3)$ & $24.4931$ \\
13 & $505.9124$ & $8.3368(-4)$ & $24.5070$ \\
14 & $505.6840$ & $6.6728(-4)$ & $24.5181$ \\
15 & $505.5002$ & $5.4237(-4)$ & $24.5270$ \\
16 & $505.3496$ & $4.4684(-4)$ & $24.5343$ \\
17 & $505.2250$ & $3.7251(-4)$ & $24.5404$ \\
18 & $505.1205$ & $3.1372(-4)$ & $24.5455$ \\
19 & $505.0322$ & $2.6694(-4)$ & $24.5498$ \\
20 & $504.9568$ & $2.2886(-4)$ & $24.5534$ \\
21 & $504.8918$ & $1.9803(-4)$ & $24.5566$ \\
30 & $504.5661$ & $6.7396(-5)$ & $24.5724$ \\
40 & $504.4299$ & $2.8420(-5)$ & $24.5791$ \\
50 & $504.3669$ & $1.4547(-5)$ & $24.5821$ \\
limit & $ 504.2549$ &        ---        & $24.5876$ \\
           \hline
      \end{tabular}
\end{table}
%
\begin{table}
        \centering
        \caption{Energies and $f$-values for the first members of a number of doubly excited $^1P^o$ Rydberg series of He. Results for the three series below the $N=2$ ionization threshold were taken from \citep{Ch97} and \citep{Li01}. Data of states converging to $N=3$-$7$ thresholds are based on Eq. (\ref{ENn}) and fits with the relation $f_n=C/n^3$. Numbers in brackets indicate powers of ten.}
       \label{t.fdsum}
       \begin{tabular}{lccc} 
       \hline
       \hline
$n$ & $E_n$[eV]$\hskip.4in f_n\hskip.2in$ & $n$ & 
                                    $E_n$[eV]$\hskip.4in f_n
       \hskip.2in$\\
       \hline
    &       ${(2,0)}_n$           &   &    ${(2,1)}_n$          \\
$2$ & $60.144$ $\quad$ $6.98(-3)$ &$3$& $62.756$ $\quad$ $2.87(-5)$ \\
$3$ & $63.654$ $\quad$ $1.14(-3)$ &$4$& $64.132$ $\quad$ $2.18(-5)$ \\
$4$ & $64.462$ $\quad$ $4.60(-4)$ &$5$& $64.655$ $\quad$ $1.11(-5)$ \\
$5$ & $64.812$ $\quad$ $2.30(-4)$ &$6$& $64.909$ $\quad$ $6.15(-6)$ \\
$6$ & $64.996$ $\quad$ $1.32(-4)$ &$7$& $65.052$ $\quad$ $3.73(-6)$ \\
$7$ & $65.105$ $\quad$ $8.25(-5)$ &   & \\
    &&&\\
    &       ${(2,-1)}_n$          &   &     ${(3,1)}_n$          \\
$3$ & $64.116$ $\quad$ $2.49(-6)$ &$3$& $69.892$ $\quad$ $1.38(-3)$ \\
$4$ & $64.646$ $\quad$ $2.22(-6)$ &$4$& $71.549$ $\quad$ $5.82(-4)$ \\
$5$ & $64.904$ $\quad$ $1.46(-6)$ &$5$& $72.152$ $\quad$ $2.99(-4)$ \\
$6$ & $64.049$ $\quad$ $9.67(-7)$ &$6$& $72.437$ $\quad$ $1.72(-4)$ \\
    &&&\\
    &       ${(4,2)}_n$           &   &     ${(5,3)}_n$          \\
$4$ & $73.729$ $\quad$ $4.36(-4)$ &$5$& $75.569$ $\quad$ $1.79(-4)$ \\
$5$ & $74.607$ $\quad$ $2.23(-4)$ &$6$& $76.088$ $\quad$ $1.03(-4)$ \\
$6$ & $74.987$ $\quad$ $1.29(-4)$ &$7$& $76.342$ $\quad$ $6.51(-5)$ \\
$7$ & $75.185$ $\quad$ $8.14(-5)$ &$8$& $76.484$ $\quad$ $4.36(-5)$ \\
    &&&\\
    &       ${(6,4)}_n$           &   &     ${(7,5)}_n$          \\
$6$ & $76.591$ $\quad$ $8.62(-5)$ &$7$& $77.217$ $\quad$ $4.65(-5)$ \\
$7$ & $76.923$ $\quad$ $5.43(-5)$ &$8$& $77.442$ $\quad$ $3.12(-5)$ \\
$8$ & $77.100$ $\quad$ $3.64(-5)$ &$9$& $77.571$ $\quad$ $2.19(-5)$ \\
$9$ & $

77.206$ $\quad$ $2.55(-5)$ &$10$&$77.651$ $\quad$ $1.60(-5)$ \\
       \hline
     \end{tabular}
\end{table}
Energies of doubly excited states of atomic helium are lying in the continuum, i.e., over the first-ionization threshold $I_1$ and below the double-ionization threshold $I_\infty=79.0052$~eV. 
In double photoexcitation from the He ground state, the dipole-selection rule allows only $^1P^o$ final states which form Rydberg series with the ``inner'' electron in a given $N$th quantum state of He$^+$, and the ``outer'' electron in $n=2,3,4,$... shells. 
Here, we adopt the state classification scheme of \citet{Li84} in the abbreviated form $(N,K)_n$ given by \citet{Zu89}. Thus, doubly photoexcited states are designed $(N,K)_n$, $K$ ranging from $N-1-T$, $N-3-T$,... to $-(N-1-T)$, with $T=0$ or 1. \footnote{The quantum numbers $K$ and $T$ were introduced by \citet{HS75} to describe the strong electron-electron correlation.}
For each $N$, there are $2N-1$ $^1P^o$ Rydberg series that converge to a same ionization threshold of He, $I_N$, which corresponds to the $N$ state of He$^+$. Some of lowest $I_N$ values are $I_2=65.404$~eV, $I_3=72.962$~eV, $I_4=75.606$~eV ,  and $I_5=76.826$~eV. Interseries interferences occur for $N\ge5$ due to overlaps of neighboring Rydberg series \citep{Do91}. 

Table \ref{t.fdsum} lists energies and $f$-values of first members of a number of  $^1P^o$ Rydberg series. Data corresponding to the three double-excitation $^1P^o$ Rydberg series below the $N=2$ ionization threshold were calculated by \citet{Ch97} and \citet{Li01} using the saddle-point complex-rotation method with B-spline functions. Table \ref{t.fdsum} also shows results of the principal series, $[N,(N-2)]_n$, converging to $N=3$-$7$ thresholds. Energies of these transitions were estimated by using a two-electron Rydberg formula \citep{Do91}
\begin{equation} \label{ENn}
E(N,n) = {\text I}_{\infty}-R\left[\frac4{N^2}+\frac1{(n-\mu_N)^2}\right],
\end{equation}
with
\begin{equation} \label{muN}
\mu_N=N-\left[\frac{2(2-\sigma)^2}{(N-\mu)^2}-\frac{4}{N^2}\right]^{-1/2},
\end{equation}
where $\mu=-0.1815$ and $\sigma=0.1587$ proceeds from a fit of Eq. (\ref{ENn}) to energies measured for Wannier ridge states $[N,(N-2)]_{n=N}$ at $N=3$-$5$. 
The $f$-values of $[N,(N-2)]_n$ states  at $N=3$-$7$ were evaluated under the assumption $f_n=C/n^3$, which is asymptotically valid for large $n$. The constant $C$ was fitted with the reduced transition probabilities obtained by \citet{Do96} for these Rydberg series. We have also used the relation $f_n=C/n^3$ for excited states $(2,K)_n$ with $n\ge7$ at $K=0,1$ and $n\ge6$ at $K=-1$.

\begin{figure}  
\centering  
\includegraphics[clip,width=250pt]{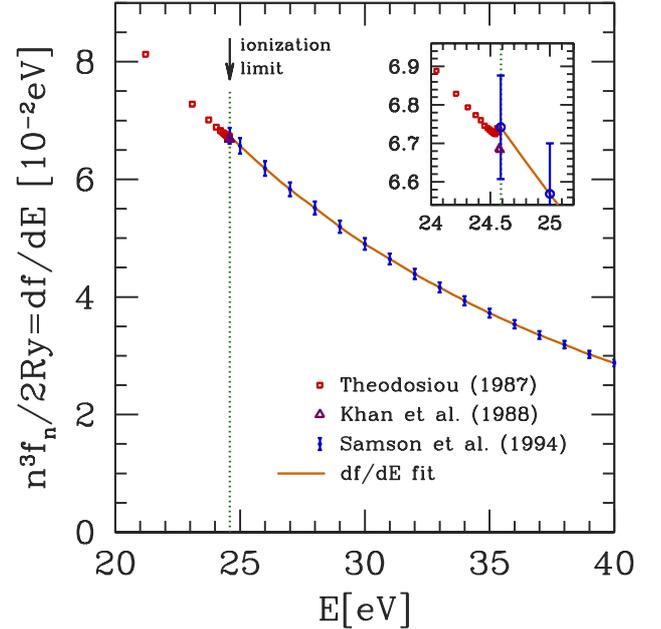}  
\caption{Oscillator strength distribution in the discrete and continuum spectra for one-electron excitation/ionization from the He ground state. The squares and triangles are, respectively, theoretical $f$-values from \citet{Th87} and \citet{Kh88}. The circles are experimental photoionization data from \citet{Sa94} with errorbars of 2\%. The solid line gives our fit to data.}
\label{f.fdis}
\end{figure}
The differential oscillator strength for transitions from the ground state to the continuum was evaluated with Eq. (\ref{dfdE}) using the photoionization cross section given by \citet{Sa94}, which corresponds to their own measurements from the ionization threshold $I_1$ to $120$~eV (with an accuracy ranging from 1 to 2\%) and recommended cross sections from $120$~eV up to $8$~keV (with an estimated uncertainty of $10$\% above $500$~eV). 
Fig. \ref{f.fdis} displays the oscillator strength distribution for single electron jumps, and shows the merging of discrete features into the continuous distribution across the ionization threshold.

The photoionization cross section of the He ground state has well-known autoionizing resonances in the $59$-$72$~eV energy region associated to double-excitation states. We have included the effects of the main resonances which are due to the $[2,0]_n$~$^1P^o$ Rydberg series. Specifically, for energies between $58$~eV and $I_2$, the cross section was represented by a product of \citet{Fa61} profiles following \citet{Fe87}
\begin{equation}\label{reson}
\sigma(E)= a(E) \prod_{n=2}^7 \frac{(q_n+\epsilon_n)^2}{1+\epsilon_n^2},
\quad (58~\text{eV}<E<I_2),
\end{equation}
with
\begin{equation}\label{epsilon1}
a(E) = 1.5\left(\frac{58}{E}\right)^{1.6}~\text{Mbarn},
\end{equation}
and
\begin{equation}\label{epsilon0}
\epsilon_n=\frac{E-E(2,n)}{\Gamma_n/2},
\end{equation}
where $\Gamma_n$ is the linewidth FWHM ($\hbar/\Gamma_n$ is the mean life of the discrete level with respect to auto-ionization), $q_n$ is the Fano-$q$ parameter that represents the ratio of the transition probabilities to a discrete state and to the continuum, and $\epsilon_n$ is the departure of the incident photon energy $E$ from a resonance energy $E(2,n)$. Values of $\Gamma_n$ and $q_n$ used in Eq. (\ref{reson}) were taken from \citet{Do96}.
\begin{figure}  
\centering  
\includegraphics[clip,width=250pt]{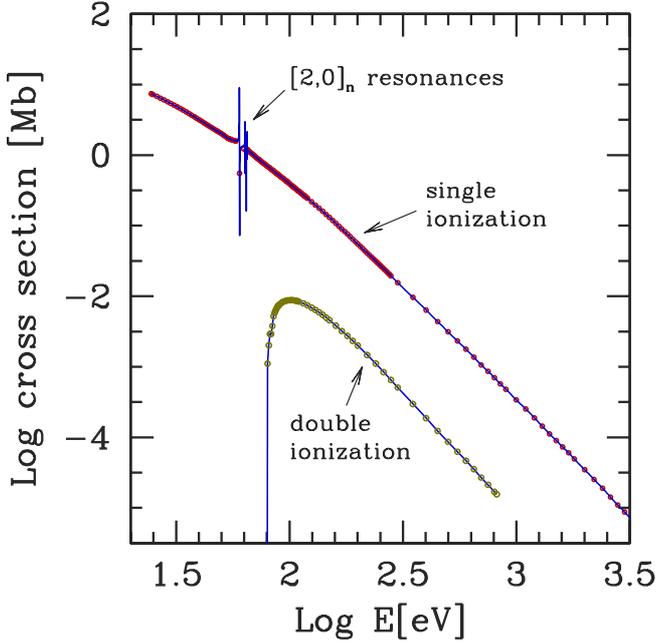}  
\caption{Symbols represent cross-sections data measured for single \citep{Sa94} and double \citep{Sa98} photoionizations. Solid lines indicate our fits. Single photoionization curve includes Fano perfiles due to resonances associated to the $[2,0]_n$~$^1P^o$ Rydberg series.}
\label{f.sec2}
\end{figure}
Double photoionization of helium has been taken from \citet{Sa98}.
Cross sections for single and double photoionizations are shown in Fig \ref{f.sec2}. The curve associated to one-electron transitions shows the resonances due to autoionizing states around 65~eV. On the other hand, double photoionizations starts at $I_\infty=79.0052$~eV and exhibits a maximum at 102~eV.

%
\begin{table}
        \centering
        \caption{Contributions to the Thomas-Reiche-Kuhn $f$-sum rule. The $f$-sum for double excited states includes the three $^1P^o$ series converging to the $N=2$ threshold and the $[N,N-2]_n$~$^1P^o$ series converging to $N=3,...,7$ thresholds.}
       \label{t.fsum}
       \begin{tabular}{llr} 
       \hline
Contribution            & $f$-sum  & Percent of total\\
       \hline
\quad$f_{n=2}$               & $0.27643$ &  $13.82$ \\
\quad$\sum_{n=3}^{21}f_n$    & $0.14837$ &   $7.42$ \\
\quad$\sum_{n>21}f_n$        & $0.00196$ &   $0.10$ \\
total discrete single jumps  & $\mathbf{0.42481}$ &  $\mathbf{21.24}$ \\
  &&\\
\quad $[2,0]_n$ series  & $0.00959$ & $0.47$ \\
\quad $[2,1]_n$ series  & $0.00008$ & $<0.01$ \\
\quad $[2,-1]_n$ series & $0.00001$ & $<0.01$ \\
\quad $[3,1]_n$ series  & $0.00287$ & $0.14$ \\
\quad $[4,2]_n$ series  & $0.00112$ & $0.06$ \\
\quad $[5,3]_n$ series  & $0.00054$ & $0.03$ \\
\quad $[6,4]_n$ series  & $0.00030$ & $0.02$ \\
\quad $[7,5]_n$ series  & $0.00018$ & $0.01$ \\
total discrete double jumps   & $\mathbf{0.01461}$ & $\mathbf{0.73}$ \\
  &&\\
continuum single jumps  & $\mathbf{1.55590}$ &  $\mathbf{77.80}$ \\
continuum double jumps  & $\mathbf{0.00724}$ &   $\mathbf{0.36}$ \\ 
\quad$\sum f$ (total)   & $\mathbf{2.00256}$ & $\mathbf{100.13}$ \\
       \hline
     \end{tabular}
\end{table}

The oscillator strengths satisfy the Thomas-Reiche-Kuhn (TRK) sum rule,
\begin{equation}\label{TRK}
\sum_n f_n +\int \frac{df}{dE} dE =\mathcal{Z},
\end{equation}
with $\mathcal{Z}$ the number of electrons ($\mathcal{Z}=2$ for helium).
The TRK rule can be used for accuracy tests in experimental and theoretical evaluations, and as an accurate absolute scale in spectrum measurements \citep{Ch91}. Here we use the TRK rule to check the precision and completeness of the oscillator strength data.
Contributions to Eq. (\ref{TRK}) from the infinite number of excited states near to an ionization threshold  (for single and double excitations), were evaluated with a Euler-Maclaurin formula \citep{AS70} taking into account the $n^{-3}$ dependence of the oscillator strength for high members of Rydberg sequences, 
\begin{equation} 
\sum_{n=n_0}^\infty f_n \approx \int_{n_0}^{\infty} 
f_n dn - \left.\frac12\frac{df_n}{dn}\right|_{n=n_0} 
+\left.\frac1{720}\frac{d^3f_n}{dn^3}\right|_{n=n_0}.
\end{equation} 

Table \ref{t.fsum} shows partial and total summations of $f$-values coming from differents sets of allowed dipole transitions from the ground state. It is noteworthly that the TRK rule is fulfilled within 0.1\%.
The dominant contribution proceeds from one-electron photoionizations ($\approx 78$\%), followed by one-electron discrete transitions ($\approx 21$\%). The double transitions to discrete and continuum states contribute 0.73\% and 0.36\% to the $f$-sum, respectively. 
Althought $f$ values from doubly excited transitions are hard to evaluate, they are involved in the accuracy of the oscillator-strength sum rule.
It is possible to appreciate that the contributions to the $f$-sum from the first six main series $[N,(N-2)]_n$ decrease faster than $N^{-3}$. Besides, secondary series $[2,1]_n$ and $[2,-1]_n$ amounts disminish at least by two and three orders relative to that of the main series $[2,0]_n$.
These results let us estimate that the omitted series of doubly-excited states have a contribution to the $f$-sum lower than $0.0005$. Therefore, the lead error in the total $f$-sum is likely due to single photoionization measurements ($1$\% of relative error of these data represents $0.8$\% of error in the total $f$-sum). Nevertheless, current $f$-sum is in much better agreement with the TRK rule than the predicted by these data errors.

\subsection{Dipole polarizability evaluations}  
\label{ss.pol} 

The dynamic dipole polarizability for the ground state of helium was calculated with the spectral distributions of atomic energy and dipole oscillator strength detailed in previous section. 
Table \ref{t.alfa} shows that our value for the static dipole polarizability is in very good agreement (within 0.005\%) with the last experimental result based on refractivity measurements with a microwave cavity \citep{Sc07}, $\alpha(0)=1.383 757[\pm13]$.
Because the decreasing dependence with the energy in the summation of Eq. ({\ref{sellmeier}), contributions from low energy transitions increase respect to the TRK rule. However, single photoionizations remain as the main source of static polarizability ($52.6$\%). Clearly, the continuum and non-resonant excitation contributions to the atomic polarizability cannot be neglected. Like TRK rule, two-electron transitions give very small contributions to the polarizability value but increase its accuracy.

%
\begin{table}
        \centering
        \caption{Static polarizability of the helium atom, showing the contributions from different radiative transitions. Last column indicates the fraction respect to the experimental value \citep{Sc07}. Numbers in brackets indicate power of ten.}
       \label{t.alfa}
       \begin{tabular}{llcr} 
       \hline
Contribution & $\alpha(0)$[a.u.] & $\alpha(0)$[cm$^3$] & [\%] \\
       \hline
        resonance line &  $0.454649$  &   $0.67372(-25)$ & $32.85$ \\
 discrete single jumps &  $0.652600$  &   $0.96705(-25)$ & $47.16$ \\
 discrete double jumps &  $0.002596$  &   $0.38472(-27)$ &  $0.19$ \\
continuum single jumps &  $0.728301$  &   $0.10792(-24)$ & $52.63$ \\
continuum double jumps &  $0.000319$  &   $0.47237(-28)$ &  $0.02$ \\
  &&\\
   total $\alpha(0)$   &  $1.383817$  &   $0.20506(-24)$ &  \\ 
   experiment        &  $1.383757[\pm13]$  &    $0.20505(-24)$ & \\
       \hline
     \end{tabular}
\end{table}

\begin{figure}  
\centering  
\includegraphics[clip,width=250pt]{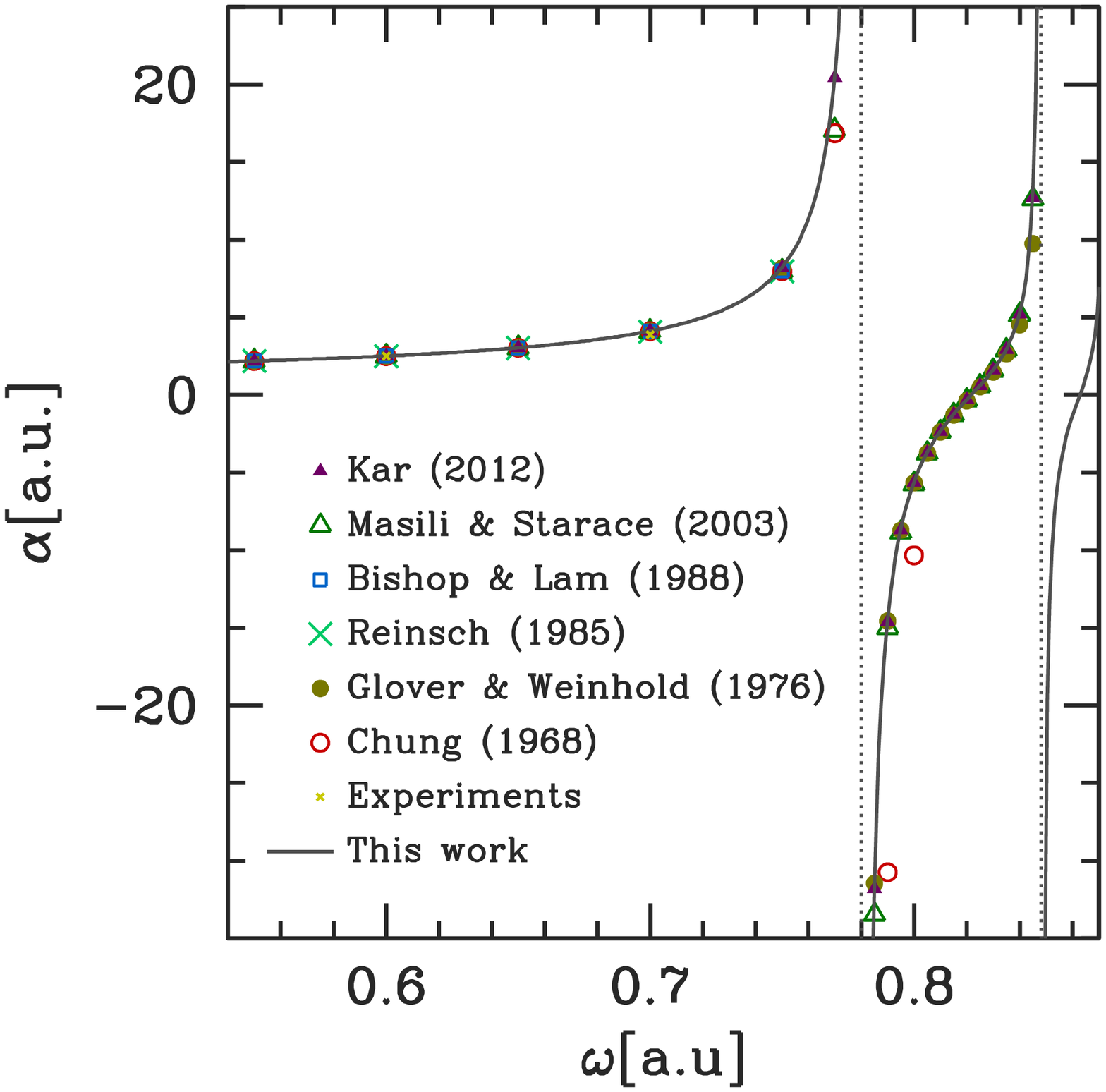}  
\caption{Dynamic atomic polarizability as a function of the angular frequency (in atomic units). Present results (solid line) are compared with those from \citet{Ka12,Ma03,Bi88,Re85,Gl76,Ch68} and experiments as fitted by Pad\'e approximants \citep{La69}.}
\label{f.alfa}
\end{figure}
Fig. \ref{f.alfa} compares the helium dynamic polarizability obtained from the present study (solid line) with available results determined by various methods (symbols), for instance, variational-perturbation method \citep{Bi88}, coupled-channel hyperspherical approach \citep{Ma03}, and pseudostate summation technique \citep{Ka12}. Those data published in reduced units were converted to atomic units using the ratio of electron to nuclear mass for the $^4$He isotope, $m_ e/M=1.37091\times10^{-4}$ \citep{Pa00}. Fig. \ref{f.alfa} includes the first two resonances (sharp antisymmetric peaks) associated to transition to intermediate $^1P$ excited states. 
One may note in the plot the good agreement of current results with the previous evaluations of \cite{Ka12} and \cite{Ma03}.

For radiation energy lower than that associated with the transition from the ground state to the lowest excited state $1s2^1P$ (i.e., $E<21.2180$~eV, $\lambda>584.334$~\AA, $\omega<0.779747$~a.u.), the dynamic polarizability can be expanded in right powers of $E$ \citep{Fc68}
\beq \label{alphaexpansion}
\alpha(E)=4\sum_{k=1}^\infty S_{2k} E^{2(k-1)}.
\eeq
With $E$ measured in Rydbergs, coefficients $S$ are given by
\beq
S_j= \sum_n \frac{f_n}{E_n^j} +\int \frac{df/dE'}{E'^j} dE'.
\eeq
Alternatively, the polarizability may be expressed as an expansion in powers of the angular frequency ($\omega=E/\hbar$)
\begin{equation} \label{alphaomega}
\alpha(\omega)=\mu_0 +\mu_1\omega^2 +\mu_2\omega^4 +\mu_3\omega^6+\dots .
\end{equation}
With $\omega$ measured in atomic units (Hartree$/\hbar$), the so-called Cauchy moments $\mu_j$ are given by\footnote{A relationship between the refractive index and the wavelength analogous to Eq. (\ref{alphaomega}) was deduced by Cauchy in 1836 \citep{KB32}.} 
\begin{equation}
\mu_j= 2^{2(j+1)}S_{2(j+1)}.
\end{equation}
Of course, the zero-order Cauchy moment is the static polarizability.
Cauchy moments may be used to construct Pad\'e approximants which are useful for fast numerical evaluations below the first excitation frequency. Besides, Pad\'e approximants are superior to Taylor series when the function contain poles as in the case of the atomic polarizability \citep{La70}. 
%
\begin{table}
        \centering
        \caption{First coefficients $S_{2(j+1)}$ in series (\ref{alphaexpansion}) and Cauchy moments $\mu_j$ of the helium static polarizability. Extra digits are displayed to avoid round off error. The last column shows the Cauchy moments computed by \citet{Pu16}. Numbers in brackets indicate power of ten.}
       \label{t.cauchy}
       \begin{tabular}{rrrl} 
       \hline
$j$ & $S_{2(j+1)}\hskip.3in$ & $\mu_j\hskip.2in$ & Puchalski et al. \\
       \hline
  0 & $3.459542(-1)$  & $1.383817$ & \hskip.05in$1.383809986408$ \\    
  1 & $9.642968(-2)$  & $1.542875$ & \hskip.05in$1.54321081882$ \\ 
  2 & $3.191798(-2)$  & $2.042751$ & \hskip.05in$2.0426550150$ \\   
  3 & $1.145032(-2)$  & $2.931283$ & \hskip.05in$2.9304069980$ \\   
  4 & $4.294122(-3)$  & $4.397181$ & \hskip.05in$4.39500532$ \\   
  5 & $1.654543(-3)$  & $6.777009$ & \hskip.05in$6.7725956$ \\   
  6 & $6.488265(-4)$  &$10.630373$ & $10.622083$ \\   
  7 & $2.575097(-4)$  &$16.876157$ & $16.86118$ \\
  8 & $1.030719(-4)$  &$27.019676$ & \quad- \\
  9 & $4.151036(-5)$  &$43.526771$ & \quad- \\
 10 & $1.679386(-5)$  &$70.438541$ & \quad- \\
       \hline
     \end{tabular}
\end{table}
Table \ref{t.cauchy} compares the Cauchy moments obtained from the present study with high accurate evaluations of \citet{Pu16}, which include relativistic and quantum electrodynamic corrections (quantum electrodynamic effects shift the polarizability value by about 0.002\%).
Both sets of $\mu_j$ values differ in the four digit or higher. 
The agreement is quite satisfactory and reveals that the oscillator-strength distribution method applied with the current data, yields reliable results for astrophysical applications.

\begin{figure}  
\centering  
\includegraphics[clip,width=250pt]{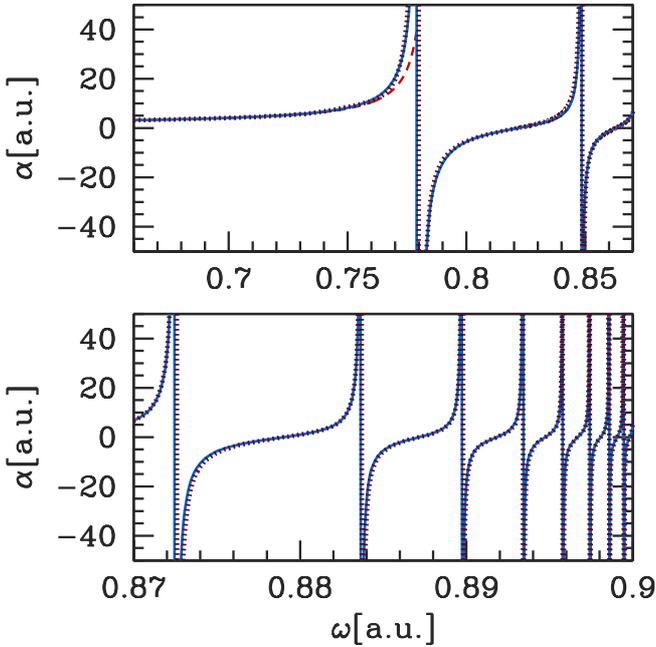}  
\caption{Dynamic atomic polarizability calculated with the oscillator-strength distribution approach (solid line) and fits  (dotted line) based on Eq. (\ref{pade}) at $\omega<0.7339$, Eq. (\ref{exp}) at $0.7339<\omega<\omega_1$, and Eq. (\ref{tan})  at $\omega_j<\omega<\omega_{j+1}$, where $\omega_j$ ($j=1,2,\dots$) are the resonance frequencies listed in Table \ref{t.tan}. The dashed line in the upper panel shows the Pade approximation given by Eq. (\ref{pade}).}
\label{f.alfa2}
\end{figure}

As shown in Fig. \ref{f.alfa2}, the dynamic polarizability (in bohr$^3$) of the helium ground state can be fitted by simple analytical expressions. At frequencies lower that the first resonance ($\omega_1=0.779748$~a.u.) 
a Pad\'e approximant is a good choice
\begin{equation}\label{pade}
\alpha(\omega) = \frac{\sum_{j=0}^3 P_j \omega^{2j}}
               {1+\sum_{j=1}^3 Q_j \omega^{2j}},
\end{equation}
with $P_0=\mu_0=1.383817$, and
\begin{eqnarray}
P_1 = -2.7631796,&& Q_1=-3.1117233,\cr
P_2 = +1.2159289,&& Q_2=+2.8718963,\cr
P_3 = -0.0188510,&& Q_3=-0.7404416.    
\end{eqnarray}
Eq. (\ref{pade}) has outstanding accuracy at low frequencies (e.g., within $0.01$\% of the exact value at $\omega<0.64$~a.u.) but get worse near 
$\omega_1$ (dashed line in Fig. \ref{f.alfa2}). For $0.7339<\omega<\omega_1$, a better performance is given by
\begin{equation}\label{exp}
\alpha(\omega) = \exp(1.6520099+22.475155 z+1.0197470\times 10^{12} z^9),
\end{equation}
with
\begin{equation}
  z = \omega-0.7282.
\end{equation}
For $\omega_j<\omega<\omega_{j+1}$ ($j=1,2,\dots$), where $\omega_j$ is the frequency of the $j$th-resonance, a trigonometric representation can be adopted with an accurate within a few percents,
\begin{equation}\label{tan}
 \alpha(\omega) = \frac{\alpha'_j}{b_j}\tan\left[b_j(\omega-\omega_{0j})\right],
\end{equation}
where
\begin{equation}
 b_j =\frac{\pi}{2(\omega_{0j}-\omega_j)},\hskip.3in  \omega<\omega_{0j},
\end{equation}
and
\begin{equation}
 b_j =\frac{\pi}{2(\omega_{j+1}-\omega_{0j})},\hskip.3in \omega>\omega_{0j}.
\end{equation}
Fitted parameters for Eq. (\ref{tan}) are listed in Table \ref{t.tan}.
%
\begin{table}
        \centering
        \caption{Coefficients used in the approximation (\ref{tan}) for the atomic dynamic polarizability at resonance region. Atomic units are used.}
       \label{t.tan}
       \begin{tabular}{rrrr} %
       \hline
$j$ & $\omega_j\quad$ & $\omega_{0j}\quad$ & $\alpha'_j\quad$ \\
       \hline
   1  &  0.779748  &  0.821611  &      186.9873\\
   2  &  0.848433  &  0.862814  &      465.6428\\
   3  &  0.872505  &  0.879035  &     1003.9436\\
   4  &  0.883667  &  0.887165  &     1728.4265\\
   5  &  0.889738  &  0.891823  &     2702.5630\\
   6  &  0.893403  &  0.894740  &     4452.0236\\
   7  &  0.895782  &  0.896694  &     6436.1707\\
   8  &  0.897415  &  0.898068  &     9005.2470\\
   9  &  0.898583  &  0.899066  &    12444.8560\\
  10  &  0.899448  &  0.899812  &    17253.3920\\
  11  &  0.900105  &  0.900386  &    21009.2542\\
  12  &  0.900617  &  0.900845  &    30141.8518\\
  13  &  0.901024  &  0.901212  &    39888.7213\\
  14  &  0.901352  &  0.901522  &    71608.2446\\
       \hline
     \end{tabular}
\end{table}
%

\subsection{Refractive index}  
\label{s.n} 

%
\begin{figure}  
\centering  
\includegraphics[clip,width=250pt]{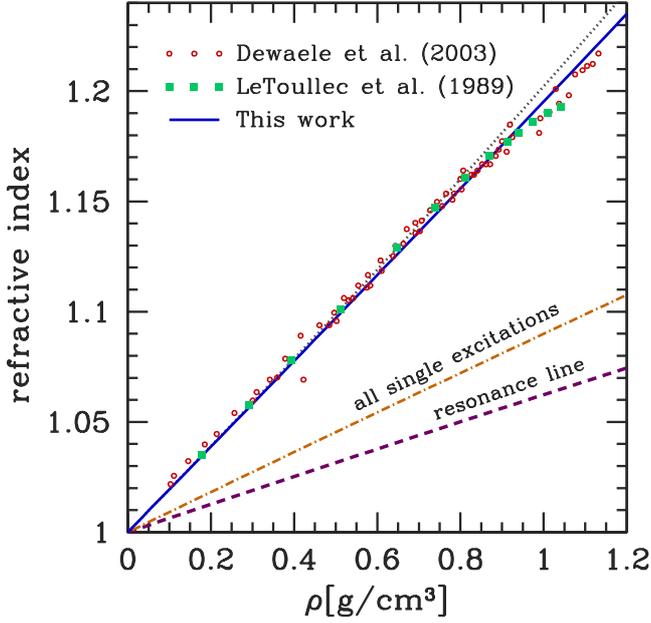}  
\caption{Refractive index for helium as a function of the mass density. Symbols represent results of measurements by \citet{Le89} (squares) and \citet{De03} (circles) at 6328~\AA. Current evaluations based on the Lorentz-Lorenz relation and full electric dipole transitions are shown by the solid line. Dotted line represents evaluations of Eq. (\ref{LL}) at first density order in the virial expansion. Dot-dashed and dotted lines indicate respectively the contributions from the resonance line $1^1S\rightarrow2^1P$ (the only contribution considered in \citet{Ig02}) and all single transitions ($1^1S\rightarrow n^1P$).}
\label{f.ref}
\end{figure}
%
Optical properties of the fluid and the atomic dipole polarizability are connected from the dispersion theory. As is well known, the refractive index $n_r$ can be determined from the Lorentz-Lorenz equation \citep{Hi54}
\begin{equation}\label{LL}
\frac{n_r^2-1}{n_r^2+2} =A_R \rho_m + B_R \rho_m^2 + ...
\end{equation}
where $\rho_{m}$ is the density in moles per unit volume, and $A_R$ is the molar polarizability or first refractivity virial coefficient, and $B_R$ the second one. The molar polarizability describes the isolated atom contribution to the refractive index and is proportional to the atomic polarizability
\begin{equation} \label{Aalpha}
A_R = \frac{4\pi}{3} N_A \alpha,
\end{equation}
with $N_A$ the Avogadro constant. The second coefficient $B_R$ accounts for the effect of two-body interactions on the refractive index. 
Accurate values of  $A_R=0.5213\pm0.0001$, $B_R=-0.068\pm0.010$ near room temperature have been measured for gaseous helium by \citet{Ac91} using a differential-interferometric technique.

\begin{figure}  
\centering  
\includegraphics[clip,width=250pt]{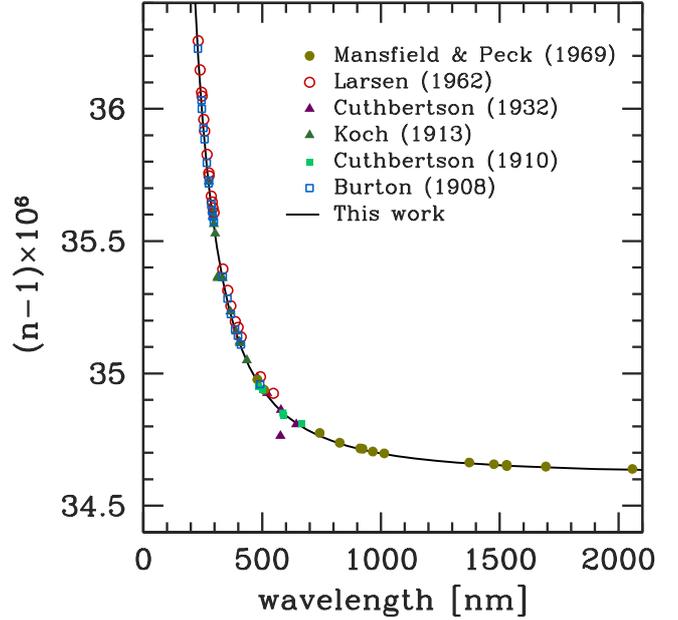}  
\caption{Refractive index for helium as a function of the wavelength. Symbols show experimental values from \citet{Ma69,La62,Cu32,Cu10,Ko13,Bu08}. Solid line shown the present evaluation for $\rho=1.788\times 10^{-4}$~g/cm$^3$, which correspond to pure helium under standard conditions (a pressure of one standard atmosphere and a temperature of 273.16~K).}
\label{f.ref_wave}
\end{figure}
The  refractive index of helium illustrated in Fig. \ref{f.ref} (solid line) is based on the $A_R$ value obtained with (\ref{sellmeier}) and (\ref{Aalpha}), and the ratio $B_R/A_R=0.1304$ derived from meassurements of \citet{Ac91}.
The figure shows an excellent agreement between current evaluations and experiments of \citet{Le89} and \citet{De03} for densities below $0.8$~g$/$cm$^3$. The introduction of the second refractivity virial coefficient $B_R$ yields a remarkably good agreement with experiments at higher densities as well. As shown in Fig. \ref{f.ref}, contributions from excited ($n^1P$, with $n>2$) and continuum states are a significant part of the total refractivity index. As the density rises over 1~g/cm$^3$, ternary and high-order particle interactions could have significant contributions to the refractive index (see Section \ref{s.cils}).

Fig. \ref{f.ref_wave} shows the helium refractive index plotted against radiation wavelength under standard pressure ($1.013250\times10^6$~dine$/$cm$^2$) and temperature ($273.16$~K). Solid line represents our results and symbols correspond to measurements in the ultraviolet \citep{La62,Bu08}, visible \citep{Cu32,Cu10,Ko13} and infrared \citep{Ma69}. The agreement is quite good.

\subsection{Scattering from free-atoms}  
\label{s.Ray0} 

\begin{figure}  
\centering  
\includegraphics[clip,width=250pt]{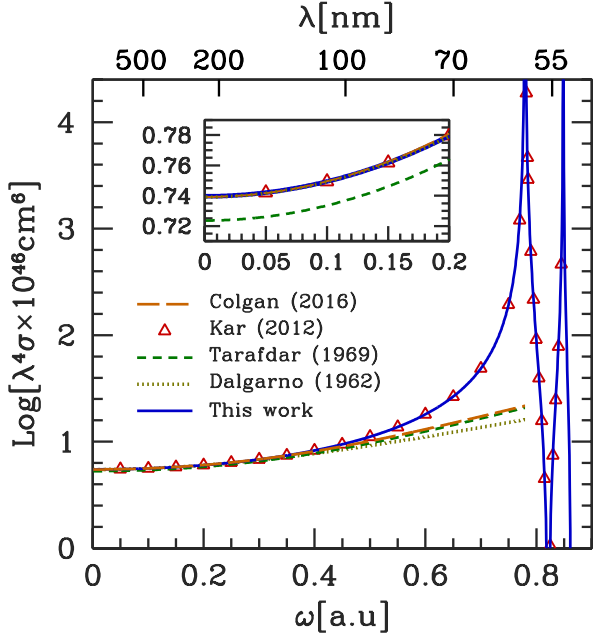}  
\caption{Rayleigh scattering cross section for free atoms including the first two resonances. The current calculation (solid line) is compared with those of \citet{Da62}, \citet{Ta69} and \citet{Co16}. Symbols represent the cross section values obtained with polarizability data from \citet{Ka12}.}
\label{f.ray}
\end{figure}
For unpolarized light, the dot product of the polarization vectors in Eq. (\ref{sigma2}) yields the usual dipole-type angular distribution, so that the integration over all directions gives
\beq \label{e1e2}
\oint \left|\mb{e}_1.\mb{e}^*_2\right|^2 \,d\Omega =8\pi/3. 
\eeq
In this case, the scattering cross section from free atoms reduces to
\beq \label{sigma0}
\sigma_0(\omega)=\frac{8\pi}{3c^4}\omega^4 \alpha(\omega)^2,
\eeq
which has the typical leading term proportional to $\omega^4$. Fig. \ref{f.ray} compares our evaluations (solid line) of the Rayleigh-scattering cross section for free helium atoms with various other theoretical approximations from \citet{Da62}, \citet{Ta69} and \citet{Co16} (dotted, dashed and long-dashed lines, respectively). \citet{Co16} evaluation is based on \citet{Da62} and \citet{DK60}, but rescaled with an updated value of the static dipole polarizability. As seen in the figure, results of \citet{Ta69} desviates from other ones at long wavelengths. Our calculation extends further to previous evaluations, and shows the appearance of first resonances associated to one-photon transitions to intermediate $^1P$ excited states. Current results are in agreement with values calculated using the polarizability values of \citet{Ka12} (symbols in the figure). With the fits given by Eqs. (\ref{pade}), (\ref{exp}) and (\ref{tan}) into expression (\ref{sigma0}) is possible to obtain the Rayleigh cross-section with a reasonable precision at $\omega<\omega_{15}=0.901620$~a.u. ($\lambda>505.35$~\AA).

\section{Density effects in scattering}  
\label{s.s} 

A basic function to describe the spatial structure of fluids is the radial distribution function $g(r)$, which represents the probability density of find a particle to a distance $r$ from a reference one. This is related to the structure function $S(k)$ througth the relations \citep{Ha06}
\begin{equation}\label{Sk} 
S(k) = 1 + n_\text{He} \hat{h}(k),
\end{equation}
and
\begin{equation}
h(r)=g(r)-1,
\end{equation}
where $n_\text{He}$ is the number density of particles in the fluid, and $\hat{h}(k)$ is the Fourier transform in three space dimension of the total correlation function $h(r)$. At real values, $\hat{h}(k)$ is given by
\beq \label{hk}
\hat{h}(k)  = \int_0^{\infty} h(r)\frac{\sin(kr)}{kr}4\pi r^2dr.
\eeq
Eq. (\ref{hk}) is evaluated by numerical quadrature taking into account at $kr<<1$ the approximation
\bea
\frac{\sin(kr)}{kr}\approx
1-\frac{(kr)^2}{6} +\frac{(kr)^4}{120} -\frac{(kr)^6}{5040}
+\frac{(kr)^8}{362880} -\frac{(kr)^{10}}{39916800}.
\eea

We obtain the pair distribution function $g(r)$ from the Monte Carlo (MC)  technique with a numerical code which we wrote following \citet{Me53}. Particles are placed within a cubic box and periodic boundary conditions are used.  For a specific configuration of mass density and temperature ($\rho,T$), the radial distribution function $g(r)$ is derived as the ratio 
\begin{equation}\label{gr}
g(r)  = \left\langle \frac{\Delta N (r)}{4\pi n r^2\Delta r} \right\rangle,
\end{equation}
where $\Delta N(r)$ is the number of particles placed at a distance from $r$ to $r + \Delta r$ from an arbitrary reference particle. 
The symbol $\langle ...\rangle$ means an average over a number $M$ of MC realizations, by taking in turn all simulation particles as the reference one. We have adopted $M=10000$ steps after 1000 ones for thermalization.  Each MC run demands several hours in a usual desk computer.

\begin{figure}  
\centering  
\includegraphics[clip,width=240pt]{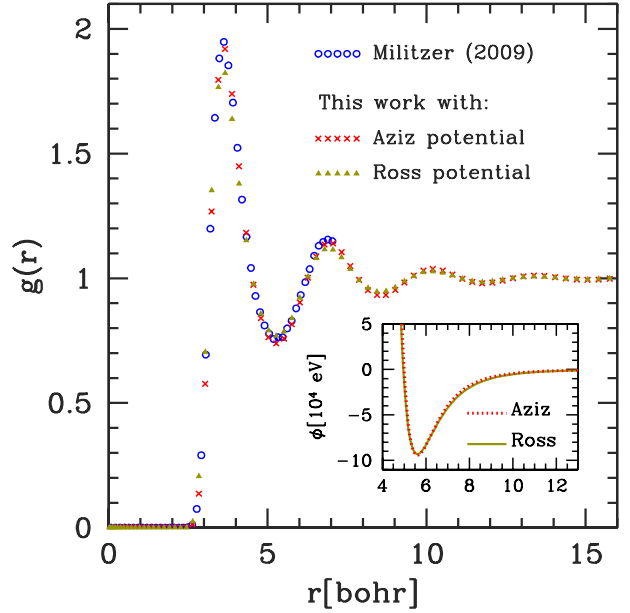}  
\caption{Comparison of the atom-atom pair distribution function from density functional molecular dynamic (DFT-MD) \citep{Mi09} with present Monte Carlo simulations based on effective interatomic potentials $\phi(r)$ from \citet{RY86} and \citet{Az95}, which are shown in the insert figure. These evaluations correspond to $T=1000$~K and $\rho=1$~g/cm$^3$.}
\label{f.militzer}
\end{figure}
The investigated systems consist of $\mathcal{N}=512$ helium atoms interacting via an effective pair potential $\phi(r)$. We have considered  analytical fits to the He-He interaction potential proposed by \citet{RY86}, and the form given in \citet{Az90} with values recommended by \citet{Az95}. The non-ideality of the gas is taken into account in both potentials. Fig. \ref{f.militzer} displays the radial distribution function obtained for fluid helium at $\rho=1$~g/cm$^3$ and $T=1000$~K using these potential laws. The interaction curves are shown in the insert figure. Short-range repulsive interactions dominate the behaviour of $g(r)$. Because the He-He attractive well is very small (depth less than $10^{-3}$~eV), this has a negligible influence on the spatial correlations. 
Aziz potential is somewhat more repulsive at short interatomic distances than the Ross \& Young one (differences can not be appreciated in the insert figure) and consequently yields slightly more pronounced oscillations in $g(r)$.
Results based on the Aziz potential compare very well with the pair distribution function obtained by \citet{Mi09} using path integral Monte Carlo technique and density functional molecular dynamics, which are accurate (but numerical expensive) approaches based on first principles. On the contrary, Ross-Young potential slightly underestimates the amplitude oscillation of the radial distribution function. The following MC simulations are based on the Aziz potential.

At very low densities, $\rho\approx 0.001$~g/cm$^3$, $g(r)$ values computed from MC simulations become noisy owing to small departures from the unity (i.e, weakly correlated particles, $h(r)\approx0$).
A perturbative treatments at first density order is appropriate in this case. For $\rho<0.005$~g/cm$^3$, we have verified that the radial distribution function is very well represented by a low-density analytical expression
\begin{equation}\label{gra}
g(r)  = \exp{[-\phi(r)/k_BT]},
\end{equation}
where $k_B$ is the Boltzmann's constant. 

\begin{figure}  
\centering  
\includegraphics[clip,width=240pt]{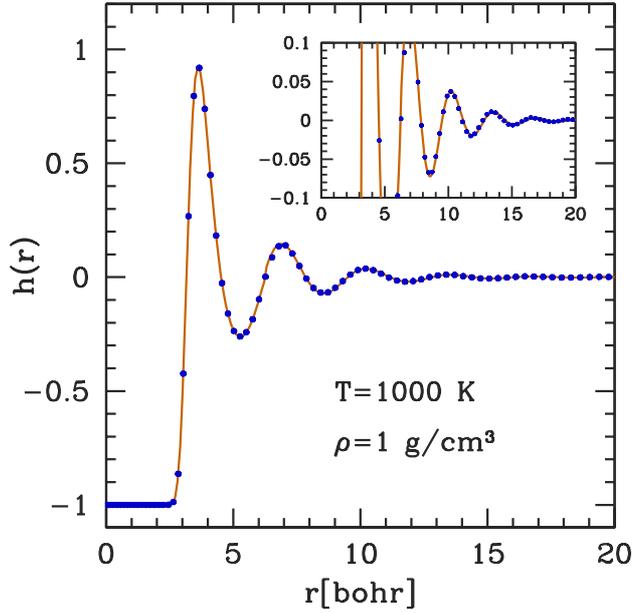} 
\caption{Correlation function for helium atoms at $T=1000$~K and $\rho=1$~g/cm$^3$. Symbols correspond to MC simulation results with the Aziz potential. The solid line is a fit based on spline cubic interpolation at $r<r_0$ and Eq. (\ref{hrfit}) for $r>r_0$, with $r_0=6.25$~bohr.}
\label{f.hr}
\end{figure}
Eq. (\ref{hk}) is solved numericaly using natural cubic splines for interpolations of $h(r)$ simulation data along radial distances. The smallest wavevector accessible from finite box simulation has magnitude $k_{min} = 2\pi/L$, where $L$ is the side length  of the cubic box. With $\rho=\mathcal{N}/L^3$, $k_{min} = 2\pi (\rho/\mathcal{N})^{1/3}$. For example, for $\mathcal{N}=512$ and $\rho=1$~g/cm$^{3}$, $k_{min} \approx 0.22$~bohr$^{-1}$. In order to obtain the small-wavenumber behaviour of $S(k)$, the large $r$ behaviour of $h(r)$ must be known with high precision. This difficulty can be partially overcome by using the long-distance asymptotic form expected for the correlation function \citep{Ki39}, 
\begin{equation}\label{hrfit}
h(r) = \frac{A}{r}e^{-\alpha r}\cos\left(\beta r +\delta\right),
\end{equation}
where the free parameters (amplitude $A$, correlation length $\alpha^{-1}$, oscillation period $\beta$, and phase shift $\delta$) are calculated by fitting of simulation data. Eq. (\ref{hrfit}) expresses that the fluctuations over large scales approaches to the zero limit value corresponding to a random distribution.
Analysis of $h(r)$ asymptotic behaviour in terms of poles of $\hat{h}(k)$ \citep{At92,He92,Ev94} confirms that a wide variety of fluids exhibits an exponentially damped oscillatory decay of structural correlations.
Fig. \ref{f.hr} shows MC values (symbols) of the total correlation function and its fit (solid line). The exponentially damped oscillatory decay is precisely represented by Eq. (\ref{hrfit}).

\begin{figure}  
\centering  
\includegraphics[clip,width=250pt]{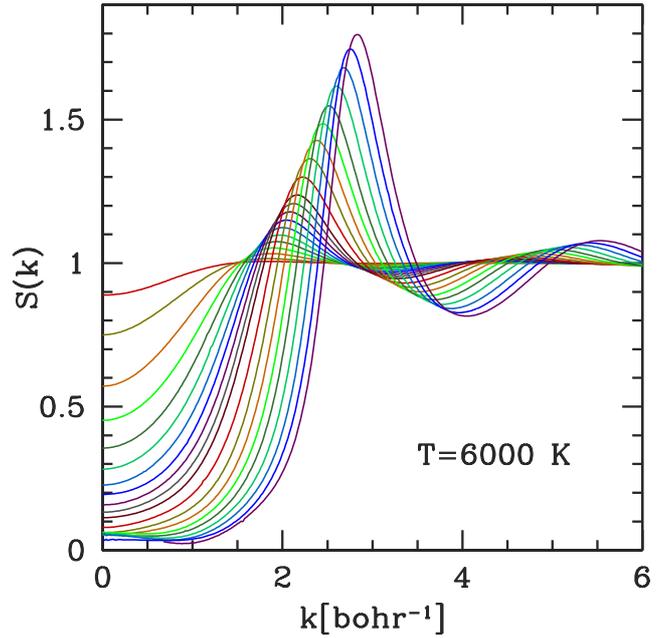}  
\caption{Static structure factor for helium at $T=6000$~K and several densities, $\rho$[g/cm$^3$]$=0.04$, $0.1$ to $1$ at $0.1$ steps, $1$ to $2$ at $0.2$ steps, and $2$ to $3$ at $0.25$ steps. Peacks increase with the fluid density.
}
\label{f.SkT6000}
\end{figure}
Fig. \ref{f.SkT6000} displays results for the static structure function at $T=6000$~K and densities between $0.04$ and $3$~g/cm$^3$. The static structure factor shows peaks which become more sharp and shifted to higher wavenumber as the density increases. Besides, $S(k)$ goes to unity at large $k$ for all densities but significantly lowers for small $k$ values when spatial correlations are present. Unfortunately, the fit given by Eq. (\ref{hrfit}) does not completely solve the precision of structure factor calculation for high density ($\rho>2$~g/cm$^3$) and low wavenumber ($k<1$~bohr$^{-1}$).

\begin{figure}  
\centering  
\includegraphics[clip,width=250pt]{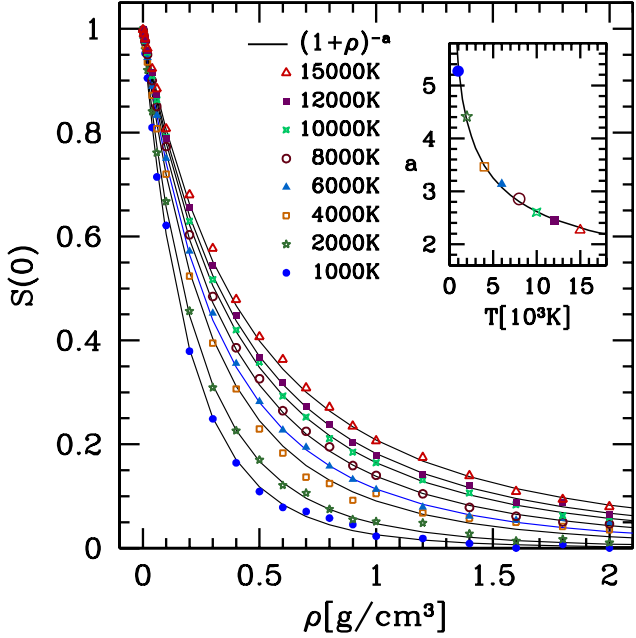}  
\caption{The structure factor at zero wavenumber as a function of the density and for different values of temperature. Curves show fits to the data with Eq. (\ref{S0empirico}), which includes a temperature-dependent parameter $a(T)$. The insert figure displays values of $a(T)$ and its fit with Eq. (\ref{S0aT}).}
\label{f.S0}
\end{figure}
As we have seen in Section \ref{tools}, the correlation effects on the scattering cross section are directly represented by the static structure function $S(k)$. For the low frequency region of interest in astrophysical applications, the moment  $k$ transferred from the photon to the atom is neglegible. Therefore, the value of $S(k)$ relevant to the Rayleigh scattering is $S(0)$, i.e., the desviation of the cross section respect to the uncorrelated systems (for which $S(k)=1,~\forall~k$) becomes proportional to $S(0)$,
\beq \label{sigma}
\sigma(\omega)=\sigma_0(\omega)S(0),
\eeq
with $\sigma_0(\omega)$ given by Eq. (\ref{sigma0}).
A decrease of the structure factor at low $k$ values (as shown in Fig. \ref{f.SkT6000}) is related to a destructive interference of the scattered field pattern and implies a reduction of the Rayleigh cross-section.

Fig. \ref{f.S0} contains $S(0)$ data derived from MC simulations for eight isotherms between 1000~K and 15000~K. More specifically, MC simulations cover the high-density range where particle correlation effects are important, while for densities lower than $0.005$~g/cm$^3$ we use the low density approximation given by Eq. (\ref{gra}). The correlation among particles increases with the density and decreases with the temperature. As is depicted in Fig. \ref{f.S0}, structure data can be well fitted by
\beq  \label{S0empirico}
S(0) = \left(1+\rho\right)^{-a(T)},
\eeq
and
\beq \label{S0aT}
a(T) = \frac{46.67685}{T^{0.3128}}.
\eeq
These fits connect smoothly to our MC data for whole $\rho-T$ points and become appropriate for opacity calculations in models of stellar atmospheres and giant planet envelopes.

A direct comparison of our results with those of \citet{Ig02} is only possible for the values $T=3240$~K and $\rho=1.23$~g/cm$^3$ reported in their fig. 2. For these conditions, they obtained a cross section correction about $0.064$ and our result is $S(0)=0.054$. Although the details of each calculation varied in terms of approach and complexity, the discrepancy can be due to the fact that \citet{Ig02} computed $S(k)$ from an approximate theoretical method, the so-called hypernetted-chain approximation, whose performance worsens for high densities \citep{Ha06}. 

\begin{figure}  
\centering  
\includegraphics[clip,width=250pt]{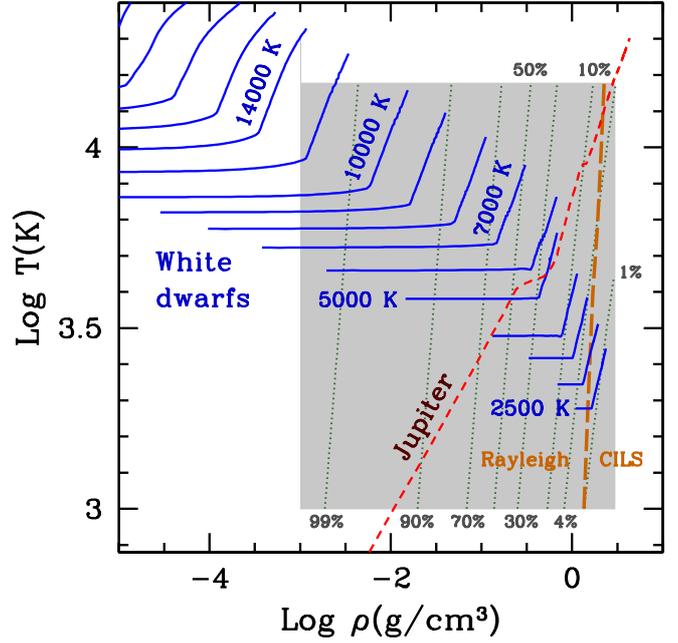}  
\caption{Density-temperature plane. The shared area shows the region studied in the present work. The contour levels (dotted curves) represent relative values of Rayleigh scattering respect to a fluid without particle correlations.
Solid lines represent temperature-density profiles of helium atmospheres of cool white dwarfs \citep{Ca17} for a surface gravity $\log g=8$ and different effective temperatures (some values are indicated on the plot).  The top and bottom of each model is located at Rosseland optical depths $\tau_{\text{Ross}}=10^{-4}$ and $100$, respectively. The short-dashed line represents the Jupiter interior model of \citet{Ne12}. The long-dashed curve divides the diagram according to the dominant scattering process, Rayleigh opacity from single atoms or collision-induced light scattering (Section \ref{s.cils}).}
\label{f.He}
\end{figure}
The physical conditions analyzed across the regime of dense helium atmospheres in cool white dwarfs and Jupiter's interior, as depicted in Fig. \ref{f.He}. Particle correlation leads to a significant reduction of the scattering cross-section for densities greater than $0.01$~g/cm$^3$. The percentage by which the total cross section is deceased depends strongly on the fluid density and more weakly on the temperature. Substantial modifications to this opacity take place in white dwarfs with surface temperature of few thousand kelvin and in deep layers of Jupiter. As indicated by dotted lines, particle correlations reduce the cross section a factor between 0.1 and 0.01 in helium atmopheres with effective temperatures cooler than 4000 K.

\subsection{Collision-induced light scattering}  
\label{s.cils} 

At high enough densities, collision-induced light scattering (CILS) yields polarized and depolarized Raman spectra, which add to the opacity caused by Rayleigh scattering  \citep{Le68,Ge74,Fr81,Bo89}. 
CILS arises from the excess of polarizability induced in atomic collisions mainly through multipolar induction (electric fields surrounding an atom due to neighbor particles) and electronic overlap induction (distorsion of electronic charges and electron correlation effects in close encounters). 
The simplest CILS spectrum is due to polarizabilities induced by binary collisions, which can be characterized by two invariants of the diatom polarizability tensor, trace ($\alpha_\text{d}$) and anisotropy ($\beta_\text{d}$),
\begin{equation}\label{dimer1}
\alpha_\text{d}=\frac13\left(\alpha_\parallel +2 \alpha_\perp\right),\quad
\beta_\text{d} = \alpha_\parallel - \alpha_\perp,
\end{equation}
where $\alpha_\parallel$ and $\alpha_\perp$ are the components of the polarizability tensor parallel and perpendicular to the dimer symmetry axis. All  quantities in Eq. (\ref{dimer1}) depend on the internuclear separation $r$ of the dimer. In this section, we estimate the relative intensities of Rayleigh and dimer CILS processes in fluid helium.

Accurate static dipole polarization of helium dimers were calculated by \citet{Ce11}. The corresponding trace and anisotropy components are shown in Fig. \ref{f.cils} (upper panel). It is useful to compare them with the classical prediction based on the dipole-induced-dipole approach
\begin{equation}
\beta_\text{DID} = \frac{6\alpha_0^2}{r^3},
\end{equation}
where $\alpha_0$ is the static polarizability of an atom. The classical induction gives a good approximation of the actual anisotropy at interatomic distances larger than 6~bohr (Fig. \ref{f.cils}), and remains below a factor 2.5 up to distances as short as 2.46~bohr (the mean interatomic distance corresponding to a density $\rho\approx 3$~g/cm$^3$). Negative values of the polarizability trace at short distances ($r\la 6.1$~bohr) are a consequence of the electronic overlap \citep{Ce11}. Since $|\alpha_\text{d}|$ remains lower than $\beta_\text{d}$ for the whole range of $r$ of interest, the analysis of collision-induced phenomenon and its significance relative to the Rayleigh scattering can be restricted to the anisotropy part. 
\begin{figure}  
\centering  
\includegraphics[clip,width=250pt]{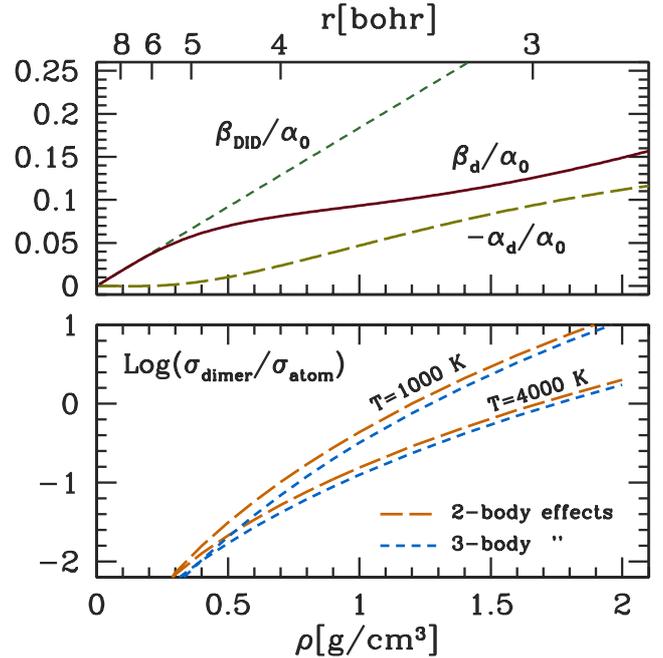}  
\caption{
{\it Upper panel:} The static trace (long-dashed line) and anisotropy (solid line) polarizabilities of helium dimer computed by \citet{Ce11} as functions of the interatomic separation ($r$, upper scale) and the density (lower scale, adopting $r$ as the mean interatomic distance in the fluid, i.e. $\rho=3m_\text{He}/4\pi r^3$). Short-dashed line represents the classical anisotropy based on the dipole-induced-dipole model. {\it Lower panel:} The ratio of cross sections of depolarized Raman scattering ($\sigma_\text{dimer}$) and Rayleigh scattering ($\sigma_\text{atom}$) taking into account two-body and three-body correlations, as a function of the fluid density at $T=1000$~K and $4000$~K.}
\label{f.cils}
\end{figure}

Expressions for the relative intensities of Rayleigh and CILS processes were derived by \citet{Ge72} and applied to argon gas, within the assumptions of static limit ($\omega=0$), dipole-induced-dipole model, and Kirkwood superposition approximation to describe three-body correlations. 
The ratio between  induced Raman and Rayleigh scattering cross-sections is predicted to be \citep{Ge72}
\begin{equation} \label{ratio}
\frac{\sigma_\text{dimer}}{\sigma_\text{atom}} =
\frac{24\pi\alpha_0^2 n_\text{He}} {5\,S(0)}  
  \left\{\int_0^\infty dr r^{-4}g(r)
+\frac{4}\pi\int_0^\infty dq P^2(q)\left[S(q)-1\right] \right\}
\end{equation}
with 
\begin{equation}
P(q)=q\left[\frac{\sin(\sigma q)}{(\sigma q)^3}
 -\frac{\cos(\sigma q)}{(\sigma q)^2}\right],
\end{equation}
$\sigma$ being the Lennard-Jones radius of an atom ($\sigma =2.6413813$~\AA~ for helium, \citet{Az95}). The term proportional to the first integral on the r.h.s. of Eq. (\ref{ratio}) corresponds to isolated pair interactions. The second integral takes into account configurations where three particles interact simultaneously in the superposition approximation.

The lower panel of Fig. \ref{f.cils} shows evaluations of Eq. (\ref{ratio}) for fluid helium using our simulation data at $T=1000$~K and $4000$~K.
The cross-section associated to CILS spectra resulting from binary collisions is roughly two orders of magnitude smaller than that of individual scattering at $\rho=0.3$~g/cm$^3$, but this difference gradually decreases at higher densities. Three-body correlations (short dashed-lines in Fig. \ref{f.cils}) yield destructive interferences (negative correlations from dipole moments induced in consecutive collisions) which reduce the many-body polarizability. 
As shown in Figs. \ref{f.He} and \ref{f.cils}, CILS becomes the dominant scattering process over $\rho\approx 1.3$~g/cm$^3$ at $T=1000$~K and from higher densities as the temperature increases ($1.7$~g/cm$^3$ at  $4000$~K, $2.1$~g/cm$^3$ at $10000$~K). These density values represent a conservative limit, since the actual anisotropy takes values lower than those provided by the DID model used in the present evaluations.


\section{Conclusions}  
\label{conc} 

We have investigated the Rayleigh scattering cross section of fluid helium under the conditions present in the envelopes of cool white dwarf stars and sub-stellar objects. In such non-relativistic conditions, the scattering cross section factorizes into an isolated-atom term depending on the atomic dipole polarizability and a collective term which accounts interference effects from the spatial distribution of scatters. On the one hand, helium polarizability evaluations has been revisited with the oscillator-strength distribution technique. The use of available transition-probability data over the full relevant spectrum of intermediate states, let us to obtain reliable dynamic polarizability values and related quantities (e.g., refractive index) in good agreement with experiments and calculations published elsewhere. On the other hand, Monte Carlo simulations of the fluid structure for eight isotherms between $1000$~K and $15000$~K and densities between 0.005~g/cm$^{-3}$ and a few g/cm$^{-3}$, let us analyze the particle correlation effects on the Rayleigh scattering. The computed corrections to the cross-section have been fitted by an analytical expression throughout the density-temperature plane, as required in hydrogen-deficient white dwarf and planetary models. In addition, we have analyzed the role of the collision-induced light scattering and found that this reaches greater intensity than the atomic Rayleigh scattering at densities higher than 1--2~g/cm$^3$ depending on the temperature.

\section*{Acknowledgements}

The author is grateful to the referee, Lothar Frommhold, for his valuable remarks which contributed to improving the quality of this work, and to Burkhard Militzer for useful comments and to share numerical data used in Fig. \ref{f.militzer}.  Valuable comments from A. Dewaele and C. Iglesias are also acknowledged. This research was supported by the Consejo Nacional de Investigaciones Cient\'ificas y T\'ecnicas (CONICET, Argentina) through Grant No. PIP 114-201101-00208 and the Ministerio de Ciencia, Tecnolog\'ia e Innovaci\'on Productiva (Argentina) through Grant No.  PICT 2016-1128.




\bibliographystyle{mnras}
\bibliography{bibHe} 

\begin{thebibliography}{}
\makeatletter
\relax
\def\mn@urlcharsother{\let\do\@makeother \do\$\do\&\do\#\do\^\do\_\do\%\do\~}
\def\mn@doi{\begingroup\mn@urlcharsother \@ifnextchar [ {\mn@doi@}
  {\mn@doi@[]}}
\def\mn@doi@[#1]#2{\def\@tempa{#1}\ifx\@tempa\@empty \href
  {http://dx.doi.org/#2} {doi:#2}\else \href {http://dx.doi.org/#2} {#1}\fi
  \endgroup}
\def\mn@eprint#1#2{\mn@eprint@#1:#2::\@nil}
\def\mn@eprint@arXiv#1{\href {http://arxiv.org/abs/#1} {{\tt arXiv:#1}}}
\def\mn@eprint@dblp#1{\href {http://dblp.uni-trier.de/rec/bibtex/#1.xml}
  {dblp:#1}}
\def\mn@eprint@#1:#2:#3:#4\@nil{\def\@tempa {#1}\def\@tempb {#2}\def\@tempc
  {#3}\ifx \@tempc \@empty \let \@tempc \@tempb \let \@tempb \@tempa \fi \ifx
  \@tempb \@empty \def\@tempb {arXiv}\fi \@ifundefined
  {mn@eprint@\@tempb}{\@tempb:\@tempc}{\expandafter \expandafter \csname
  mn@eprint@\@tempb\endcsname \expandafter{\@tempc}}}

\bibitem[\protect\citeauthoryear{{Abramowitz} \& {Stegun}}{{Abramowitz} \&
  {Stegun}}{1970}]{AS70}
{Abramowitz} M.,  {Stegun} I.~A.,  1970, {Handbook of mathematical functions :
  with formulas, graphs, and mathematical tables}.
Dover Publications, INC., New York

\bibitem[\protect\citeauthoryear{{Achtermann}, {Magnus}  \&
  {Bose}}{{Achtermann} et~al.}{1991}]{Ac91}
{Achtermann} H.~J.,  {Magnus} G.,   {Bose} T.~K.,  1991, \mn@doi [\jcp]
  {10.1063/1.460478}, \href {http://adsabs.harvard.edu/abs/1991JChPh..94.5669A}
  {94, 5669}

\bibitem[\protect\citeauthoryear{{Attard}, {Ursenbach}  \& {Patey}}{{Attard}
  et~al.}{1992}]{At92}
{Attard} P.,  {Ursenbach} C.~P.,   {Patey} G.~N.,  1992, \mn@doi [\pra]
  {10.1103/PhysRevA.45.7621}, \href
  {http://adsabs.harvard.edu/abs/1992PhRvA..45.7621A} {45, 7621}

\bibitem[\protect\citeauthoryear{{Aziz} \& {Slaman}}{{Aziz} \&
  {Slaman}}{1990}]{Az90}
{Aziz} R.~A.,  {Slaman} M.~J.,  1990, \mn@doi [Metrologia]
  {10.1088/0026-1394/27/4/005}, \href
  {http://adsabs.harvard.edu/abs/1990Metro..27..211A} {27, 211}

\bibitem[\protect\citeauthoryear{{Aziz}, {Janzen}  \& {Moldover}}{{Aziz}
  et~al.}{1995}]{Az95}
{Aziz} R.~A.,  {Janzen} A.~R.,   {Moldover} M.~R.,  1995, \mn@doi [Phys. Rev.
  Lett.] {10.1103/PhysRevLett.74.1586}, \href
  {http://adsabs.harvard.edu/abs/1995PhRvL..74.1586A} {74, 1586}

\bibitem[\protect\citeauthoryear{{Ben-Jaffel} \& {Abbes}}{{Ben-Jaffel} \&
  {Abbes}}{2015}]{Be15}
{Ben-Jaffel} L.,  {Abbes} I.,  2015, in Journal of Physics Conference Series.
  p. 012003 (\mn@eprint {arXiv} {1410.0672}),
  \mn@doi{10.1088/1742-6596/577/1/012003}

\bibitem[\protect\citeauthoryear{{Benneke} \& {Seager}}{{Benneke} \&
  {Seager}}{2012}]{Be12}
{Benneke} B.,  {Seager} S.,  2012, \mn@doi [\apj]
  {10.1088/0004-637X/753/2/100}, \href
  {http://adsabs.harvard.edu/abs/2012ApJ...753..100B} {753, 100}

\bibitem[\protect\citeauthoryear{{B{\'e}tr{\'e}mieux}}{{B{\'e}tr{\'e}mieux}}{2016}]{Be16}
{B{\'e}tr{\'e}mieux} Y.,  2016, \mn@doi [\mnras] {10.1093/mnras/stv2955}, \href
  {http://adsabs.harvard.edu/abs/2016MNRAS.456.4051B} {456, 4051}

\bibitem[\protect\citeauthoryear{{Bishop} \& {Lam}}{{Bishop} \&
  {Lam}}{1988}]{Bi88}
{Bishop} D.~M.,  {Lam} B.,  1988, \mn@doi [\pra] {10.1103/PhysRevA.37.464},
  \href {http://adsabs.harvard.edu/abs/1988PhRvA..37..464B} {37, 464}

\bibitem[\protect\citeauthoryear{{Bonin} \& {Kresin}}{{Bonin} \&
  {Kresin}}{1997}]{Bo97}
{Bonin} K.~D.,  {Kresin} V.~V.,  1997, {Electric-dipole polarizabilities of
  atoms, molecules, and clusters}.
World Scientific Publishing

\bibitem[\protect\citeauthoryear{{Borysow} \& {Frommhold}}{{Borysow} \&
  {Frommhold}}{1989}]{Bo89}
{Borysow} A.,  {Frommhold} L.,  1989, Advances in Chemical Physics, 75, 439

\bibitem[\protect\citeauthoryear{{Burton}}{{Burton}}{1908}]{Bu08}
{Burton} J.,  1908, Proc. Roy. Soc. (London) Series A, 80, 390

\bibitem[\protect\citeauthoryear{{Camisassa}, {Althaus}, {Rohrmann},
  {Garc{\'{\i}}a-Berro}, {Torres}, {C{\'o}rsico}  \& {Wachlin}}{{Camisassa}
  et~al.}{2017}]{Ca17}
{Camisassa} M.~E.,  {Althaus} L.~G.,  {Rohrmann} R.~D.,  {Garc{\'{\i}}a-Berro}
  E.,  {Torres} S.,  {C{\'o}rsico} A.~H.,   {Wachlin} F.~C.,  2017, \mn@doi
  [\apj] {10.3847/1538-4357/aa6797}, \href
  {http://adsabs.harvard.edu/abs/2017ApJ...839...11C} {839, 11}

\bibitem[\protect\citeauthoryear{{Carlson} \& {Judge}}{{Carlson} \&
  {Judge}}{1974}]{Ca74}
{Carlson} R.~W.,  {Judge} D.~L.,  1974, \mn@doi [\jgr]
  {10.1029/JA079i025p03623}, \href
  {http://adsabs.harvard.edu/abs/1974JGR....79.3623C} {79, 3623}

\bibitem[\protect\citeauthoryear{{Cencek}, {Komasa}  \& {Szalewicz}}{{Cencek}
  et~al.}{2011}]{Ce11}
{Cencek} W.,  {Komasa} J.,   {Szalewicz} K.,  2011, \mn@doi [\jcp]
  {10.1063/1.3603968}, \href
  {http://adsabs.harvard.edu/abs/2011JChPh.135a4301C} {135, 014301}

\bibitem[\protect\citeauthoryear{{Chan}, {Cooper}  \& {Brion}}{{Chan}
  et~al.}{1991}]{Ch91}
{Chan} W.~F.,  {Cooper} G.,   {Brion} C.~E.,  1991, \mn@doi [\pra]
  {10.1103/PhysRevA.44.186}, \href
  {http://adsabs.harvard.edu/abs/1991PhRvA..44..186C} {44, 186}

\bibitem[\protect\citeauthoryear{{Chen}}{{Chen}}{1997}]{Ch97}
{Chen} M.-K.,  1997, \mn@doi [\pra] {10.1103/PhysRevA.56.4537}, \href
  {http://adsabs.harvard.edu/abs/1997PhRvA..56.4537C} {56, 4537}

\bibitem[\protect\citeauthoryear{{Chen} \& {Kotlarchyk}}{{Chen} \&
  {Kotlarchyk}}{2007}]{ck07}
{Chen} S.-H.,  {Kotlarchyk} M.,  2007, {Interactions of photons and neutrons
  with matter}.
World Scientific

\bibitem[\protect\citeauthoryear{{Chung}}{{Chung}}{1968}]{Ch68}
{Chung} K.~T.,  1968, \mn@doi [Physical Review] {10.1103/PhysRev.166.1}, \href
  {http://adsabs.harvard.edu/abs/1968PhRv..166....1C} {166, 1}

\bibitem[\protect\citeauthoryear{{Colgan} et~al.,}{{Colgan}
  et~al.}{2016}]{Co16}
{Colgan} J.,  et~al., 2016, \apj, \href
  {http://adsabs.harvard.edu/abs/2016ApJ...817..116C} {817, 116}

\bibitem[\protect\citeauthoryear{{Cuthbertson} \& {Cuthbertson}}{{Cuthbertson}
  \& {Cuthbertson}}{1910}]{Cu10}
{Cuthbertson} C.,  {Cuthbertson} M.,  1910, \mn@doi [Proceedings of the Royal
  Society of London Series A] {10.1098/rspa.1910.0052}, \href
  {http://adsabs.harvard.edu/abs/1910RSPSA..84...13C} {84, 13}

\bibitem[\protect\citeauthoryear{{Cuthbertson} \& {Cuthbertson}}{{Cuthbertson}
  \& {Cuthbertson}}{1932}]{Cu32}
{Cuthbertson} C.,  {Cuthbertson} M.,  1932, \mn@doi [Proceedings of the Royal
  Society of London Series A] {10.1098/rspa.1932.0019}, \href
  {http://adsabs.harvard.edu/abs/1932RSPSA.135...40C} {135, 40}

\bibitem[\protect\citeauthoryear{{Dalgarno}}{{Dalgarno}}{1962a}]{Da62}
{Dalgarno} A.,  1962a, {Spectral Reflectivity of the Earth's Atmosphere III:
  The Scattering of Light by Atomic Systems}.
Geophysical Corporation of America Rep.

\bibitem[\protect\citeauthoryear{{Dalgarno}}{{Dalgarno}}{1962b}]{Da62a}
{Dalgarno} A.,  1962b, \mn@doi [Advances in Physics]
  {10.1080/00018736200101302}, \href
  {http://adsabs.harvard.edu/abs/1962AdPhy..11..281D} {11, 281}

\bibitem[\protect\citeauthoryear{{Dalgarno} \& {Kingston}}{{Dalgarno} \&
  {Kingston}}{1960}]{DK60}
{Dalgarno} A.,  {Kingston} A.~E.,  1960, \mn@doi [Proceedings of the Royal
  Society of London Series A] {10.1098/rspa.1960.0237}, \href
  {http://adsabs.harvard.edu/abs/1960RSPSA.259..424D} {259, 424}

\bibitem[\protect\citeauthoryear{{Dalgarno} \& {Lynn}}{{Dalgarno} \&
  {Lynn}}{1957}]{Da57}
{Dalgarno} A.,  {Lynn} N.,  1957, \mn@doi [Proceedings of the Physical Society
  A] {10.1088/0370-1298/70/11/303}, \href
  {http://adsabs.harvard.edu/abs/1957PPSA...70..802D} {70, 802}

\bibitem[\protect\citeauthoryear{{Dewaele}, {Eggert}, {Loubeyre}  \& {Le
  Toullec}}{{Dewaele} et~al.}{2003}]{De03}
{Dewaele} A.,  {Eggert} J.~H.,  {Loubeyre} P.,   {Le Toullec} R.,  2003,
  \mn@doi [\prb] {10.1103/PhysRevB.67.094112}, \href
  {http://adsabs.harvard.edu/abs/2003PhRvB..67i4112D} {67, 094112}

\bibitem[\protect\citeauthoryear{{Domke} et~al.,}{{Domke} et~al.}{1991}]{Do91}
{Domke} M.,  et~al., 1991, \mn@doi [Physical Review Letters]
  {10.1103/PhysRevLett.66.1306}, \href
  {http://adsabs.harvard.edu/abs/1991PhRvL..66.1306D} {66, 1306}

\bibitem[\protect\citeauthoryear{{Domke}, {Schulz}, {Remmers}, {Kaindl}  \&
  {Wintgen}}{{Domke} et~al.}{1996}]{Do96}
{Domke} M.,  {Schulz} K.,  {Remmers} G.,  {Kaindl} G.,   {Wintgen} D.,  1996,
  \mn@doi [Phys. Rev. A] {10.1103/PhysRevA.53.1424}, \href
  {http://adsabs.harvard.edu/abs/1996PhRvA..53.1424D} {53, 1424}

\bibitem[\protect\citeauthoryear{{Evans}, {Leote de Carvalho}, {Henderson}  \&
  {Hoyle}}{{Evans} et~al.}{1994}]{Ev94}
{Evans} R.,  {Leote de Carvalho} R.~J.~F.,  {Henderson} J.~R.,   {Hoyle} D.~C.,
   1994, \mn@doi [\jcp] {10.1063/1.466920}, \href
  {http://adsabs.harvard.edu/abs/1994JChPh.100..591E} {100, 591}

\bibitem[\protect\citeauthoryear{{Fano}}{{Fano}}{1961}]{Fa61}
{Fano} U.,  1961, \mn@doi [Physical Review] {10.1103/PhysRev.124.1866}, \href
  {http://adsabs.harvard.edu/abs/1961PhRv..124.1866F} {124, 1866}

\bibitem[\protect\citeauthoryear{{Fano} \& {Cooper}}{{Fano} \&
  {Cooper}}{1968}]{Fc68}
{Fano} U.,  {Cooper} J.~W.,  1968, \mn@doi [Reviews of Modern Physics]
  {10.1103/RevModPhys.40.441}, \href
  {http://adsabs.harvard.edu/abs/1968RvMP...40..441F} {40, 441}

\bibitem[\protect\citeauthoryear{{Fernley}, {Seaton}  \& {Taylor}}{{Fernley}
  et~al.}{1987}]{Fe87}
{Fernley} J.~A.,  {Seaton} M.~J.,   {Taylor} K.~T.,  1987, \mn@doi [Journal of
  Physics B Atomic Molecular Physics] {10.1088/0022-3700/20/23/032}, \href
  {http://adsabs.harvard.edu/abs/1987JPhB...20.6457F} {20, 6457}

\bibitem[\protect\citeauthoryear{{Frommhold}}{{Frommhold}}{1981}]{Fr81}
{Frommhold} L.,  1981, Advances in Chemical Physics, 46, 1

\bibitem[\protect\citeauthoryear{{Garc{\'{\i}}a-Berro} \&
  {Oswalt}}{{Garc{\'{\i}}a-Berro} \& {Oswalt}}{2016}]{Ga16}
{Garc{\'{\i}}a-Berro} E.,  {Oswalt} T.~D.,  2016, \mn@doi [\nar]
  {10.1016/j.newar.2016.08.001}, \href
  {http://adsabs.harvard.edu/abs/2016NewAR..72....1G} {72, 1}

\bibitem[\protect\citeauthoryear{{Gelbart}}{{Gelbart}}{1972}]{Ge72}
{Gelbart} W.~M.,  1972, \mn@doi [\jcp] {10.1063/1.1678301}, \href
  {http://adsabs.harvard.edu/abs/1972JChPh..57..699G} {57, 699}

\bibitem[\protect\citeauthoryear{{Gelbart}}{{Gelbart}}{1974}]{Ge74}
{Gelbart} W.~M.,  1974, Advances in Chemical Physics, 26, 1

\bibitem[\protect\citeauthoryear{{Glover} \& {Weinhold}}{{Glover} \&
  {Weinhold}}{1976}]{Gl76}
{Glover} R.~M.,  {Weinhold} F.,  1976, \mn@doi [\jcp] {10.1063/1.432967}, \href
  {http://adsabs.harvard.edu/abs/1976JChPh..65.4913G} {65, 4913}

\bibitem[\protect\citeauthoryear{{Hansen} \& {McDonald}}{{Hansen} \&
  {McDonald}}{2006}]{Ha06}
{Hansen} J.-P.,  {McDonald} I.~R.,  2006, {Theory of Simple Liquids}.
Academic Press

\bibitem[\protect\citeauthoryear{{Hargreaves}}{{Hargreaves}}{1929}]{Ha29}
{Hargreaves} J.,  1929, \mn@doi [Proc. of the Cambridge Philosophical Soc.]
  {10.1017/S0305004100018594}, \href
  {http://adsabs.harvard.edu/abs/1929PCPS...25...75H} {25, 75}

\bibitem[\protect\citeauthoryear{{Hartree}}{{Hartree}}{1928}]{Ha28}
{Hartree} D.~R.,  1928, \mn@doi [Proc. of the Cambridge Philosophical Soc.]
  {10.1017/S0305004100015954}, \href
  {http://adsabs.harvard.edu/abs/1928PCPS...24..426H} {24, 426}

\bibitem[\protect\citeauthoryear{{Henderson} \& {Sabeur}}{{Henderson} \&
  {Sabeur}}{1992}]{He92}
{Henderson} J.~R.,  {Sabeur} Z.~A.,  1992, \mn@doi [\jcp] {10.1063/1.463652},
  \href {http://adsabs.harvard.edu/abs/1992JChPh..97.6750H} {97, 6750}

\bibitem[\protect\citeauthoryear{{Herrick} \& {Sinanoglu}}{{Herrick} \&
  {Sinanoglu}}{1975}]{HS75}
{Herrick} D.~R.,  {Sinanoglu} O.,  1975, \mn@doi [\pra]
  {10.1103/PhysRevA.11.97}, \href
  {http://adsabs.harvard.edu/abs/1975PhRvA..11...97H} {11, 97}

\bibitem[\protect\citeauthoryear{{Hirschfelder}, {Curtiss}  \&
  {Bird}}{{Hirschfelder} et~al.}{1954}]{Hi54}
{Hirschfelder} J.~O.,  {Curtiss} C.~F.,   {Bird} R.~B.,  1954, {Molecular
  Theory of Gases and Liquids}.
John Wiley and Sons

\bibitem[\protect\citeauthoryear{{Howe} \& {Burrows}}{{Howe} \&
  {Burrows}}{2012}]{Ho12}
{Howe} A.~R.,  {Burrows} A.~S.,  2012, \mn@doi [\apj]
  {10.1088/0004-637X/756/2/176}, \href
  {http://adsabs.harvard.edu/abs/2012ApJ...756..176H} {756, 176}

\bibitem[\protect\citeauthoryear{{Iglesias}, {Rogers}  \& {Saumon}}{{Iglesias}
  et~al.}{2002}]{Ig02}
{Iglesias} C.~A.,  {Rogers} F.~J.,   {Saumon} D.,  2002, \mn@doi [\apjl]
  {10.1086/340689}, \href {http://adsabs.harvard.edu/abs/2002ApJ...569L.111I}
  {569, L111}

\bibitem[\protect\citeauthoryear{{Isliker}, {Nussbaumer}  \& {Vogel}}{{Isliker}
  et~al.}{1989}]{Is89}
{Isliker} H.,  {Nussbaumer} H.,   {Vogel} M.,  1989, \aap, \href
  {http://adsabs.harvard.edu/abs/1989A%26A...219..271I} {219, 271}

\bibitem[\protect\citeauthoryear{{Kar}}{{Kar}}{2012}]{Ka12}
{Kar} S.,  2012, \mn@doi [\pra] {10.1103/PhysRevA.86.062516}, \href
  {http://adsabs.harvard.edu/abs/2012PhRvA..86f2516K} {86, 062516}

\bibitem[\protect\citeauthoryear{{Khan}, {Khandelwal}  \& {Wilson}}{{Khan}
  et~al.}{1988}]{Kh88}
{Khan} F.,  {Khandelwal} G.~S.,   {Wilson} J.~W.,  1988, \mn@doi [\apj]
  {10.1086/166394}, \href {http://adsabs.harvard.edu/abs/1988ApJ...329..493K}
  {329, 493}

\bibitem[\protect\citeauthoryear{{Kirkwood}}{{Kirkwood}}{1939}]{Ki39}
{Kirkwood} J.~G.,  1939, \mn@doi [\jcp] {10.1063/1.1750344}, \href
  {http://adsabs.harvard.edu/abs/1939JChPh...7..919K} {7, 919}

\bibitem[\protect\citeauthoryear{{Koch}}{{Koch}}{1913}]{Ko13}
{Koch} J.,  1913, Arkiv Mat. Astron. Fysik, 9, 6

\bibitem[\protect\citeauthoryear{{Korff} \& {Breit}}{{Korff} \&
  {Breit}}{1932}]{KB32}
{Korff} S.~A.,  {Breit} G.,  1932, \mn@doi [Reviews of Modern Physics]
  {10.1103/RevModPhys.4.471}, \href
  {http://adsabs.harvard.edu/abs/1932RvMP....4..471K} {4, 471}

\bibitem[\protect\citeauthoryear{{Langhoff} \& {Karplus}}{{Langhoff} \&
  {Karplus}}{1969}]{La69}
{Langhoff} P.~W.,  {Karplus} M.,  1969, Journal of the Optical Society of
  America (1917-1983), \href
  {http://adsabs.harvard.edu/abs/1969JOSA...59..863L} {59, 863}

\bibitem[\protect\citeauthoryear{{Langhoff} \& {Karplus}}{{Langhoff} \&
  {Karplus}}{1970}]{La70}
{Langhoff} P.~W.,  {Karplus} M.,  1970, {``Application of Pad\'e Approximants
  to Dispersion Force and Optical Polarizability Computations'', in The Pad\'e
  Approximant in Theoretical Physics, ed. G.A. Baker, Jr., and J.L. Gammel, pp.
  41-97}.
Academic

\bibitem[\protect\citeauthoryear{{Larsen}}{{Larsen}}{1962}]{La62}
{Larsen} R.,  1962, {in Zahlenwerte and Funktionen aus Physik, Chemie,
  Astronomie, Geophysik und Technik, Vol. II, part 8, Berlin, p. 6-82}.
Springer

\bibitem[\protect\citeauthoryear{{Le Toullec}, {Loubeyre}  \& {Pinceaux}}{{Le
  Toullec} et~al.}{1989}]{Le89}
{Le Toullec} R.,  {Loubeyre} P.,   {Pinceaux} J.-P.,  1989, \mn@doi [\prb]
  {10.1103/PhysRevB.40.2368}, \href
  {http://adsabs.harvard.edu/abs/1989PhRvB..40.2368L} {40, 2368}

\bibitem[\protect\citeauthoryear{{Lecavelier Des Etangs}, {Pont},
  {Vidal-Madjar}  \& {Sing}}{{Lecavelier Des Etangs} et~al.}{2008}]{Le08}
{Lecavelier Des Etangs} A.,  {Pont} F.,  {Vidal-Madjar} A.,   {Sing} D.,  2008,
  \mn@doi [\aap] {10.1051/0004-6361:200809388}, \href
  {http://adsabs.harvard.edu/abs/2008A%26A...481L..83L} {481, L83}

\bibitem[\protect\citeauthoryear{{Levine} \& {Birnbaum}}{{Levine} \&
  {Birnbaum}}{1968}]{Le68}
{Levine} H.~B.,  {Birnbaum} G.,  1968, \mn@doi [Physical Review Letters]
  {10.1103/PhysRevLett.20.439}, \href
  {http://adsabs.harvard.edu/abs/1968PhRvL..20..439L} {20, 439}

\bibitem[\protect\citeauthoryear{{Lewis}}{{Lewis}}{2013}]{Le13}
{Lewis} A.,  2013, \mn@doi [\jcap] {10.1088/1475-7516/2013/08/053}, \href
  {http://adsabs.harvard.edu/abs/2013JCAP...08..053L} {8, 053}

\bibitem[\protect\citeauthoryear{{Lin}}{{Lin}}{1984}]{Li84}
{Lin} C.~D.,  1984, \mn@doi [\pra] {10.1103/PhysRevA.29.1019}, \href
  {http://adsabs.harvard.edu/abs/1984PhRvA..29.1019L} {29, 1019}

\bibitem[\protect\citeauthoryear{{Liu}, {Chen}  \& {Lin}}{{Liu}
  et~al.}{2001}]{Li01}
{Liu} C.-N.,  {Chen} M.-K.,   {Lin} C.~D.,  2001, \mn@doi [\pra]
  {10.1103/PhysRevA.64.010501}, \href
  {http://adsabs.harvard.edu/abs/2001PhRvA..64a0501L} {64, 010501}

\bibitem[\protect\citeauthoryear{{Loudon}}{{Loudon}}{2000}]{Lo00}
{Loudon} R.,  2000, {The quantum theory of light}.
Oxford University Press

\bibitem[\protect\citeauthoryear{{Mansfield} \& {Peck}}{{Mansfield} \&
  {Peck}}{1969}]{Ma69}
{Mansfield} C.~R.,  {Peck} E.~R.,  1969, Journal of the Optical Society of
  America (1917-1983), \href
  {http://adsabs.harvard.edu/abs/1969JOSA...59..199M} {59, 199}

\bibitem[\protect\citeauthoryear{{Martin}}{{Martin}}{1973}]{Ma73}
{Martin} W.~C.,  1973, \mn@doi [J. of Phys. and Chem. Reference Data]
  {10.1063/1.3253119}, \href
  {http://adsabs.harvard.edu/abs/1973JPCRD...2..257M} {2, 257}

\bibitem[\protect\citeauthoryear{{Masili} \& {Starace}}{{Masili} \&
  {Starace}}{2003}]{Ma03}
{Masili} M.,  {Starace} A.,  2003, \mn@doi [\pra] {10.1103/PhysRevA.68.012508},
  \href {http://adsabs.harvard.edu/abs/2003PhRvA..68a2508M} {68}

\bibitem[\protect\citeauthoryear{{Mattarelli}, {Montagna}  \&
  {Verrocchio}}{{Mattarelli} et~al.}{2007}]{Ma07}
{Mattarelli} M.,  {Montagna} M.,   {Verrocchio} P.,  2007, \mn@doi [Applied
  Physics Letters] {10.1063/1.2768642}, \href
  {http://adsabs.harvard.edu/abs/2007ApPhL..91f1911M} {91, 061911}

\bibitem[\protect\citeauthoryear{{Metropolis}, {Rosenbluth}, {Rosenbluth},
  {Teller}  \& {Teller}}{{Metropolis} et~al.}{1953}]{Me53}
{Metropolis} N.,  {Rosenbluth} A.~W.,  {Rosenbluth} M.~N.,  {Teller} A.~H.,
  {Teller} E.,  1953, \mn@doi [\jcp] {10.1063/1.1699114}, \href
  {http://adsabs.harvard.edu/abs/1953JChPh..21.1087M} {21, 1087}

\bibitem[\protect\citeauthoryear{{Militzer}}{{Militzer}}{2009}]{Mi09}
{Militzer} B.,  2009, \mn@doi [Journal of Physics A Mathematical General]
  {10.1088/1751-8113/42/21/214001}, \href
  {http://adsabs.harvard.edu/abs/2009JPhA...42u4001M} {42, 214001}

\bibitem[\protect\citeauthoryear{{Mitroy}, {Safronova}  \& {Clark}}{{Mitroy}
  et~al.}{2010}]{Mi10}
{Mitroy} J.,  {Safronova} M.~S.,   {Clark} C.~W.,  2010, \mn@doi [Journal of
  Physics B Atomic Molecular Physics] {10.1088/0953-4075/43/20/202001}, \href
  {http://adsabs.harvard.edu/abs/2010JPhB...43t2001M} {43, 202001}

\bibitem[\protect\citeauthoryear{{Nettelmann}, {Becker}, {Holst}  \&
  {Redmer}}{{Nettelmann} et~al.}{2012}]{Ne12}
{Nettelmann} N.,  {Becker} A.,  {Holst} B.,   {Redmer} R.,  2012, \mn@doi
  [\apj] {10.1088/0004-637X/750/1/52}, \href
  {http://adsabs.harvard.edu/abs/2012ApJ...750...52N} {750, 52}

\bibitem[\protect\citeauthoryear{{Pachucki} \& {Sapirstein}}{{Pachucki} \&
  {Sapirstein}}{2000}]{Pa00}
{Pachucki} K.,  {Sapirstein} J.,  2000, \mn@doi [Journal of Physics B Atomic
  Molecular Physics] {10.1088/0953-4075/33/23/303}, \href
  {http://adsabs.harvard.edu/abs/2000JPhB...33.5297P} {33, 5297}

\bibitem[\protect\citeauthoryear{{Parkinson}, {Stewart}, {Wong}, {Yung}  \&
  {Ajello}}{{Parkinson} et~al.}{2006}]{Pa06}
{Parkinson} C.~D.,  {Stewart} A.~I.~F.,  {Wong} A.~S.,  {Yung} Y.~L.,
  {Ajello} J.~M.,  2006, \mn@doi [Journal of Geophysical Research (Planets)]
  {10.1029/2005JE002539}, \href
  {http://adsabs.harvard.edu/abs/2006JGRE..111.2002P} {111, E02002}

\bibitem[\protect\citeauthoryear{{Pendrill}}{{Pendrill}}{1996}]{Pe96}
{Pendrill} L.~R.,  1996, \mn@doi [Journal of Physics B Atomic Molecular
  Physics] {10.1088/0953-4075/29/16/007}, \href
  {http://adsabs.harvard.edu/abs/1996JPhB...29.3581P} {29, 3581}

\bibitem[\protect\citeauthoryear{{Plagemann}, {R{\"u}ter}, {Bornath}, {Shihab},
  {Desjarlais}, {Fortmann}, {Glenzer}  \& {Redmer}}{{Plagemann}
  et~al.}{2015}]{Pl15}
{Plagemann} K.-U.,  {R{\"u}ter} H.~R.,  {Bornath} T.,  {Shihab} M.,
  {Desjarlais} M.~P.,  {Fortmann} C.,  {Glenzer} S.~H.,   {Redmer} R.,  2015,
  \mn@doi [Phys. Rev. E] {10.1103/PhysRevE.92.013103}, \href
  {http://adsabs.harvard.edu/abs/2015PhRvE..92a3103P} {92, 013103}

\bibitem[\protect\citeauthoryear{{Puchalski}, {Piszczatowski}, {Komasa},
  {Jeziorski}  \& {Szalewicz}}{{Puchalski} et~al.}{2016}]{Pu16}
{Puchalski} M.,  {Piszczatowski} K.,  {Komasa} J.,  {Jeziorski} B.,
  {Szalewicz} K.,  2016, \mn@doi [\pra] {10.1103/PhysRevA.93.032515}, \href
  {http://adsabs.harvard.edu/abs/2016PhRvA..93c2515P} {93, 032515}

\bibitem[\protect\citeauthoryear{{Reinsch}}{{Reinsch}}{1985}]{Re85}
{Reinsch} E.-A.,  1985, \mn@doi [\jcp] {10.1063/1.449657}, \href
  {http://adsabs.harvard.edu/abs/1985JChPh..83.5784R} {83, 5784}

\bibitem[\protect\citeauthoryear{{Ross} \& {Young}}{{Ross} \&
  {Young}}{1986}]{RY86}
{Ross} M.,  {Young} D.~A.,  1986, \mn@doi [Physics Letters A]
  {10.1016/0375-9601(86)90752-8}, \href
  {http://adsabs.harvard.edu/abs/1986PhLA..118..463R} {118, 463}

\bibitem[\protect\citeauthoryear{{Sakurai}}{{Sakurai}}{1967}]{Sa67}
{Sakurai} J.~J.,  1967, {Advanced quantum mechanics, Addinson and Wesley,
  Reading, Mass.}.
Addinson and Wesley

\bibitem[\protect\citeauthoryear{{Samson}, {He}, {Yin}  \& {Haddad}}{{Samson}
  et~al.}{1994}]{Sa94}
{Samson} J.~A.~R.,  {He} Z.~X.,  {Yin} L.,   {Haddad} G.~N.,  1994, \mn@doi [J.
  of Phys. B At. Mol. Phys.] {10.1088/0953-4075/27/5/008}, \href
  {http://adsabs.harvard.edu/abs/1994JPhB...27..887S} {27, 887}

\bibitem[\protect\citeauthoryear{{Samson}, {Stolte}, {He}, {Cutler}, {Lu}  \&
  {Bartlett}}{{Samson} et~al.}{1998}]{Sa98}
{Samson} J.~A.~R.,  {Stolte} W.~C.,  {He} Z.-X.,  {Cutler} J.~N.,  {Lu} Y.,
  {Bartlett} R.~J.,  1998, \mn@doi [Phys. Rev. A] {10.1103/PhysRevA.57.1906},
  \href {http://adsabs.harvard.edu/abs/1998PhRvA..57.1906S} {57, 1906}

\bibitem[\protect\citeauthoryear{{Schmidt}, {Gavioso}, {May}  \&
  {Moldover}}{{Schmidt} et~al.}{2007}]{Sc07}
{Schmidt} J.~W.,  {Gavioso} R.~M.,  {May} E.~F.,   {Moldover} M.~R.,  2007,
  \mn@doi [Physical Review Letters] {10.1103/PhysRevLett.98.254504}, \href
  {http://adsabs.harvard.edu/abs/2007PhRvL..98y4504S} {98, 254504}

\bibitem[\protect\citeauthoryear{{Schuelke}}{{Schuelke}}{2007}]{Sch07}
{Schuelke} W.,  2007, {Electron Dynamics by Inelastic X-ray Scattering, Oxford
  University Press}.
Oxford University Press

\bibitem[\protect\citeauthoryear{{Scopigno}, {Ruocco}  \& {Sette}}{{Scopigno}
  et~al.}{2005}]{Sc05}
{Scopigno} T.,  {Ruocco} G.,   {Sette} F.,  2005, \mn@doi [Reviews of Modern
  Physics] {10.1103/RevModPhys.77.881}, \href
  {http://adsabs.harvard.edu/abs/2005RvMP...77..881S} {77, 881}

\bibitem[\protect\citeauthoryear{{Sellmeier}}{{Sellmeier}}{1871}]{Se71}
{Sellmeier} W.,  1871, Annalen der Physik, 219, 272

\bibitem[\protect\citeauthoryear{{Sturm}}{{Sturm}}{1993}]{St93}
{Sturm} K.,  1993, \mn@doi [Zeitschrift Naturforschung Teil A]
  {10.1515/zna-1993-1-244}, \href
  {http://adsabs.harvard.edu/abs/1993ZNatA..48..233S} {48, 233}

\bibitem[\protect\citeauthoryear{{Tarafdar} \& {Vardya}}{{Tarafdar} \&
  {Vardya}}{1969}]{Ta69}
{Tarafdar} S.~P.,  {Vardya} M.~S.,  1969, \mn@doi [\mnras]
  {10.1093/mnras/145.2.171}, \href
  {http://adsabs.harvard.edu/abs/1969MNRAS.145..171T} {145, 171}

\bibitem[\protect\citeauthoryear{{Theodosiou}}{{Theodosiou}}{1987}]{Th87}
{Theodosiou} C.~E.,  1987, \mn@doi [Atomic Data and Nuclear Data Tables]
  {10.1016/0092-640X(87)90017-9}, \href
  {http://adsabs.harvard.edu/abs/1987ADNDT..36...97T} {36, 97}

\bibitem[\protect\citeauthoryear{{Van Hove}}{{Van Hove}}{1954}]{vH54}
{Van Hove} L.,  1954, \mn@doi [Physical Review] {10.1103/PhysRev.95.249}, \href
  {http://adsabs.harvard.edu/abs/1954PhRv...95..249V} {95, 249}

\bibitem[\protect\citeauthoryear{{Vinti}}{{Vinti}}{1933}]{Vi33}
{Vinti} J.~P.,  1933, \mn@doi [Physical Review] {10.1103/PhysRev.44.524}, \href
  {http://adsabs.harvard.edu/abs/1933PhRv...44..524V} {44, 524}

\bibitem[\protect\citeauthoryear{{Wheeler}}{{Wheeler}}{1933}]{Wh33}
{Wheeler} J.~A.,  1933, \mn@doi [Physical Review] {10.1103/PhysRev.43.258},
  \href {http://adsabs.harvard.edu/abs/1933PhRv...43..258W} {43, 258}

\bibitem[\protect\citeauthoryear{{Yu}, {Spergel}  \& {Ostriker}}{{Yu}
  et~al.}{2001}]{Yu01}
{Yu} Q.,  {Spergel} D.~N.,   {Ostriker} J.~P.,  2001, \mn@doi [\apj]
  {10.1086/322482}, \href {http://adsabs.harvard.edu/abs/2001ApJ...558...23Y}
  {558, 23}

\bibitem[\protect\citeauthoryear{{Zubek}, {King}, {Rutter}  \& {Read}}{{Zubek}
  et~al.}{1989}]{Zu89}
{Zubek} M.,  {King} G.~C.,  {Rutter} P.~M.,   {Read} F.~H.,  1989, \mn@doi
  [Journal of Physics B Atomic Molecular Physics]
  {10.1088/0953-4075/22/21/007}, \href
  {http://adsabs.harvard.edu/abs/1989JPhB...22.3411Z} {22, 3411}

\makeatother
\end{thebibliography}



\bsp	
\label{lastpage}
\end{document}